\documentclass[prd,aps,twocolumn,nofootinbib,superscriptaddress,eqsecnum,floatfix,preprintnumbers,amsmath,amssymb,
nofootinbib,longbibliography]{revtex4-2}

\usepackage{amssymb,amsmath}
\usepackage{epsfig}
\usepackage[dvipsnames]{xcolor}
\usepackage[utf8]{inputenc}
\usepackage{stmaryrd}
\usepackage{mathrsfs}
\usepackage{mathalfa}
\usepackage{accents}
\usepackage[normalem]{ulem}
\usepackage{graphicx}

\newcommand{\pa}{\partial}

\newcommand{\be}{\begin{equation}}
\newcommand{\ee}{\end{equation}}
\newcommand{\bea}{\begin{eqnarray}}
\newcommand{\eea}{\end{eqnarray}}
\newcommand{\ba}{\begin{equation}\begin{aligned}}
\newcommand{\ea}{\end{aligned}\end{equation}}

\newcommand{\beg}{\begin{gather*}}
\newcommand{\eng}{\end{gather*}}
\newcommand{\hh}{,\hspace{0.5cm}}
\newcommand{\hhh}{,\hspace{0.2cm}}

\newcommand{\n}[1]{\label{#1}}

\newcommand{\CAL}{\mathcal}

\newcommand{\ts}[1]{{\boldsymbol{#1}}}

\def\XXint#1#2#3{{\setbox0=\hbox{$#1{#2#3}{\int}$ }
\vcenter{\hbox{$#2#3$ }}\kern-.6\wd0}}

\usepackage[makeroom]{cancel}
\usepackage[caption=false]{subfig}
\usepackage[colorlinks=true,
            citecolor=green,
            linkcolor=red,
            filecolor=cyan,
            urlcolor=magenta,
            backref=false]{hyperref}
\begin{document}

\title{Spinoptics in the Kerr Spacetime: Polarized Wave Scattering}

\author{Valeri P. Frolov}%
\email[]{vfrolov@ualberta.ca}
\affiliation{Theoretical Physics Institute, Department of Physics,
University of Alberta,\\
Edmonton, Alberta, T6G 2E1, Canada
}

\author{Alex Koek}%
\email[]{koek@ualberta.ca}
\affiliation{Theoretical Physics Institute, Department of Physics,
University of Alberta,\\
Edmonton, Alberta, T6G 2E1, Canada
}

%\today

\begin{abstract} \label{abstract}
We study propagation of high-frequency electromagnetic and gravitational waves in the gravitational field of a rotating black hole. Due to the interaction of the spin of the field with the spacetime curvature, the standard geometric optics approximation that is used for obtaining the approximate high-frequency solutions of the wave equation should be modified. The corresponding modified spinoptics equations show that the worldline of the spinning massless particle is still null, but no longer a geodesic. We demonstrate that using the hidden symmetries of the Kerr metric one can obtain the corresponding spinoptics equations in the leading order of a $1/\omega$ expansion in an explicit form. We focus on the case of the spinning massless fields scattering in the region near the equatorial plane.  We demonstrate that the asymptotic planes of the corresponding null ray's motion are slightly tilted. We study this effect and its dependence on the spin of the black hole.

 \hfill {\scriptsize Alberta Thy 2-25}
\end{abstract}

\maketitle

\section{Introduction} \label{s1}

In this paper, we study the propagation of high-frequency electromagnetic and gravitational waves in the gravitational field of a rotating black hole. In the standard geometric optics approximation, both Maxwell's equations and the equations for gravitational waves in a curved spacetime reduce to studying a dynamical system with the Hamiltonian of the system describing motion of massless particles.
Such particles move along null geodesics, and the polarization vectors describing the spins of the particles are parallel propagated (see e.g. \cite{MTW}).

The interaction of the spin of a massless particle with the spacetime curvature modifies its motion. A corresponding worldline is still null, but not a geodesic. To describe this effect, a modification of the geometric optics approximation, known as the spinoptics approximation, was developed.
Spin-optical effects connected with the spin of a rotating black hole were described and studied in \cite{Frolov:2011mh,Frolov:2012zn,Yoo:2012vv}.
The covariant spinoptics equations for high-frequency electromagnetic waves were formulated in
\cite{Oancea:2019pgm,Oancea:2020khc,
Frolov:2020uhn,Dahal:2022gop}. Similar spinoptics equations for high-frequency gravitational waves in a curved spacetime were considered in \cite{Yamamoto:2017gla,Andersson:2020gsj,Dahal:2021qel,Kubota:2023dlz}.

The starting point of the spinoptics approximation is similar to that of geometric optics. Namely, one searches for an approximate high frequency solution in the following form
\be \n{ANSATZ}
\begin{split}
&A_{\mu}=\Re \Big[
a m_{\mu} \exp(i\omega S)\big] \hhh
\mbox{for EM waves}\, ,\\
&h_{\mu\nu}=\Re \Big[
a m_{\mu}m_{\nu} \exp(i\omega S)\big] \hhh
\mbox{for grav. waves}\, .
\end{split}
\ee
Here $\omega$ is the frequency parameter, $a$ is a real amplitude, $S$ is an eikonal function, and $m^{\mu}$ is a normalized complex null polarization vector.   Such solutions describe circularly polarized waves. These waves are characterized by fixed helicities. Changing the polarization vector $m^{\mu}$ to its complex-conjugated form transforms one circular polarization to the opposite one. To obtain spinoptics equations, one proceeds as follows
\begin{itemize}
\item Use the high-frequency $\omega$ ansatz \eqref{ANSATZ} for circularly polarized waves, making sure to distinguish between helicities, and get the high-frequency expansion.
\item Single out helicity-sensitive terms in the high-frequency expansion. \item Include $1/\omega$ helicity-dependent terms into the eikonal equation.
\end{itemize}

More recently, it was demonstrated that the spinoptics equations can be obtained from an effective action \cite{Frolov:2024ebe,Frolov:2024qow}.
In this approach, one substitutes the ansatz \eqref{ANSATZ} for the high-frequency circular polarized waves directly into the action describing the field.
Keeping the first two terms of its high-frequency expansion one obtains a reduced action. This action is a functional of the amplitude, complex null polarization vector, and eikonal function. It was shown that by varying this functional over these variables one obtains the spinoptics equations.

The solutions for the spinoptics equations in the Schwarzschild spacetime were obtained and analyzed in \cite{Dahal:2021qel,Murk:2024qgj,Frolov:2024olb}.
The main new spinoptical effect described in these papers for circularly polarized photons and gravitons passing near the black hole is the following. At a far distance from the black hole, both the incoming and outgoing orbits of the massless particles are planar. However, these in- and out-planes are tilted with respect to each other.
The spinoptic equations in the Kerr spacetime were recently discussed by Dahal \cite{Dahal:2023ncl}. This author used solutions for parallel propagated vectors along null geodesics in the Kerr metric
\cite{MARCK_1,Kubiznak:2008zs,FRO_ZEL,LIVING}
to present the spinoptic equations in an explicit form and analyzed solutions of these equations in the leading order of the parameter $M/r$.

In this paper, we study the spinoptics equations in the Kerr spacetime. We generalize the results of \cite{Dahal:2023ncl} in two ways. First, we obtain an explicit analytic expression for the helicity-dependent ``driving force" acting on the null rays due to the spin-curvature interaction, not only on the equatorial plane but outside of it as well. Second, for the null rays in the vicinity of the equatorial plane, we use the method of perturbations to integrate the spinoptical equations and obtain the tilting angle describing the deflection of the asymptotic orbit plane for outgoing rays with respect to the original orbit plane for incoming rays. We study the dependence of this angle on the spin parameter of the rotating black hole for both types of the ray's trajectories, namely prograde and retrograde rays.
We demonstrate that in the absence of rotation of the black hole, the obtained results reproduce the results presented earlier for the Schwarzschild black hole in
\cite{Murk:2024qgj,Frolov:2024olb}. The new result is the study of how the spin of the rotating black hole modifies the tilting angle for photon and graviton scattering by the black hole.

The paper is organized as follows.
In Sec.~\ref{s2} we collect useful formulas for the Kerr metric and generators of its hidden symmetries. In Sec.~\ref{s3} we construct a tetrad associated with the congruence of geodesic null rays in the Kerr geometry. Geodesic null rays and their associated tetrads in the equatorial plane are discussed in Sec.~\ref{s4}. The spinoptics equations in the Kerr spacetime, both inside and outside of the equatorial plane, are derived in Sec.~\ref{s5}. In Sec.~\ref{s6}, we discuss the scattering of high-frequency circularly polarized electromagnetic and gravitational waves propagating in the vicinity of the equatorial plane of a rotating black hole. Sec.~\ref{s7}, contains summary of the obtained results and their discussion. There are three appendices that contain additional material, which is used in the main part of the paper.

Throughout this paper, we use the sign conventions of the book \cite{MTW} and geometric units of $c=G=1$. We also denote four-dimensional objects, such as four-dimensional vectors and tensors, by boldface symbols.

\section{The Kerr metric and its symmetries} \label{s2}

\subsection{Dimensionless form of the Kerr metric} \label{s2a}

The Kerr metric in Boyer-Lindquist coordinates is
\be \n{metricDim}
\begin{split}
&dS^2=-\Big(1-\frac{2Mr}{r^2+a^2\cos^2\theta}\Big)dt^2+\frac{r^2+a^2\cos^2\theta}{r^2 -2Mr + a^2}dr^2\\
&-\frac{4Mra\sin^2\theta}{r^2+a^2\cos^2\theta}dt d\phi
+( r^2+a^2\cos^2\theta) d\theta^2\\
&+\Big(r^2+a^2+\frac{2Mra^2}{r^2+a^2\cos^2\theta}\sin^2\theta\Big)\sin^2\theta d\phi^2\, .
\end{split}
\ee
Here $M$ is the mass of the black hole and $a$ is its spin parameter. We choose $a$ to be positive and require it to satisfy the condition $0\le a\le M$.

The mass $M$ of the black hole determines a length scale. It is convenient to use it to define dimensionless parameters
\footnote{
In the absence of rotation, that is for the Schwarzschild metric, one often uses the gravitational radius $r_g=2M$ instead of $M$ in order to obtain dimensionless quantities.
}
\be \n{DIM}
\rho=r/M\hhh \tau=t/M\hhh \alpha=a/M\,,
\ee
and rewrite the metric in the dimensionless form
\be
dS^2=M^2 ds^2\, ,
\ee
where
\ba \n{metric}
&ds^2=-\Big(1-\frac{2\rho}{\Sigma}\Big)d\tau^2+\frac{\Sigma}{\Delta}d\rho^2-\frac{4\rho\alpha\sin^2\theta}{\Sigma}d\tau d\phi\\
&\hspace{25pt}+\Sigma d\theta^2+\Big(\rho^2+\alpha^2+\frac{2\rho\alpha^2}{\Sigma}\sin^2\theta\Big)\sin^2\theta d\phi^2\,,\\[8pt]
&\Sigma = \rho^2+\alpha^2\cos^2\theta\hhh\Delta = \rho^2 -2\rho + \alpha^2\hhh
\sqrt{-g}=\Sigma\sin\theta
\,.
\ea

The larger root of the equation $\Delta=0$ determines the position of the outer horizon
\be
\rho_h=1+\sqrt{1-\alpha^2}\, .
\ee
The black hole's outer ergosphere is determined by finding the largest root of the equation $g_{\tau\tau}=0$, and is located at
\be
\rho_e=1+\sqrt{1-\alpha^2\cos^2\theta}\, .
\ee

\subsection{ZAMO frames} \label{s2b}

There exists a convenient frame of reference in the Kerr spacetime. This frame is an orthonormal tetrad tied to the coordinates $\rho$ and $\theta$. It has the form
\be \n{EEEE}
\begin{split}
E_{(\tau)}&=E_{(\tau)}^{\mu}\pa_{\mu}=\sqrt{\dfrac{A}{\Delta\Sigma}}\big(
\dfrac{\pa}{\pa \tau}+\Omega \dfrac{\pa}{\pa \phi}
\big)\, ,\\
E_{(\phi)}&=E_{(\phi)}^{\mu}\pa_{\mu}=\sqrt{\dfrac{\Sigma}{A}}\dfrac{1}{\sin\theta}
\dfrac{\pa}{\pa \phi}\, ,\\
E_{(\rho)}&=E_{(\rho)}^{\mu}\pa_{\mu}=
\sqrt{\dfrac{\Delta}{\Sigma}}\dfrac{\pa}{\pa \rho}\, ,\\
E_{(\theta)}&=E_{(\theta)}^{\mu}\pa_{\mu}=
\sqrt{\dfrac{1}{\Sigma}}\dfrac{\pa}{\pa \theta}\, ,
\end{split}
\ee
where
\be
A = (\rho^2+\alpha^2)^2-\Delta \alpha^2\sin^2\theta\,.
\ee
The world line of an observer that is at rest in this frame is orthogonal to a $\tau$=constant hypersurface. This local observer rotates with respect to infinity with angular velocity
\be
\Omega=\dfrac{2\alpha\rho}{A}\, ,
\ee
and has zero angular momentum \cite{Bardeen_1,Bardeen_2,Bardeen}.
The frame \eqref{EEEE} associated with this zero angular momentum observer (ZAMO) is called a ZAMO frame.

\subsection{Explicit and hidden symmetries} \label{s2c}

The Kerr metric $ds^2$ admits two Killing vectors
\be \n{KillingVecs}
\boldsymbol{\xi} = \partial_{\tau}\hspace{0.5cm}\text{and}\hspace{0.5cm} \boldsymbol{\zeta} = \partial_{\phi}\,.
\ee
It is also invariant under the reflections
$\theta\to \pi-\theta$, and $\{\tau\to -\tau, \phi\to -\phi\}$.

A zero-mass free particle in a curved spacetime moves along a null geodesic.
Consider the null geodesic $x^{\mu}=x^{\mu}(\lambda)$, and
let $l^{\mu}$ be its null tangent vector
\be \n{intl}
\dfrac{dx^{\mu}}{d\lambda}=l^{\mu}\, .
\ee
Let $\lambda$ be a dimensionless affine parameter. Then one has
\be \n{geodl}
l^{\nu}l^{\mu}_{\ ;\nu}=0\, .
\ee
For geodesic motion in the Kerr spacetime there exist three conserved quantities: energy, projection of the angular momentum in the direction of the black hole's spin, and mass. For massless particles, which we consider in this paper, the mass vanishes and one has
\be \n{EL}
E=-\xi^{\mu}l_{\mu}\hh
L_z=\zeta^{\mu}l_{\mu}\hh
g_{\mu\nu}l^{\mu}l^{\nu}=0\, .
\ee
Let us note that an affine parameter is defined up to a  transformation
$\lambda \to A\lambda +B$, where $A$ and $B$ are constants. We use this freedom and put $A=E$ and $B=0$. As a result the relations \eqref{EL} take the form
\be \n{EL1}
1=-\xi^{\mu}l_{\mu}\hh
\ell_z=\zeta^{\mu}l_{\mu}\hh
g_{\mu\nu}l^{\mu}l^{\nu}=0\, ,
\ee
where $\ell_z=L_z/E$ is the dimensionless impact parameter. In the rest of the paper we shall use the affine parameter that gives us these dimensionless normalizations of our conserved quantities.

In 1968 Carter demonstrated that the geodesic equations in the Kerr metric are completely integrable \cite{Carter:68}. According to Liouville's theorem, complete integrability of an $n$-dimensional dynamical system requires the existence of $n$ independent commuting integrals of motion (see e.g. \cite{Arnold:1989}).
Carter showed that, in addition to the three integrals of motion given by \eqref{EL1}, there exists a fourth integral of motion, which received the name Carter's constant. This integral of motion has the form
\be\n{QQ}
Q=K_{\mu\nu} l^{\mu}l^{\nu}\, ,
\ee
where $K_{\mu\nu}$ is the Killing tensor obeying the equations
\be
K_{\mu\nu}=K_{(\mu\nu)}\hh K_{(\mu\nu ;\lambda)}=0\, .
\ee
Later Penrose and Floyd \cite{PENROSE} showed that this tensor is the ``square" of a rank two antisymmetric tensor $k_{\mu\nu}$
\be\n{KKK}
K_{\mu\nu} = k_{\mu}^{\hspace{5pt}\lambda}k_{\nu\lambda}\,.
\ee
The tensor $\ts{k}$ is called a Killing-Yano tensor, and it satisfies the equation
\be\n{kkk}
\nabla_{(\lambda}k_{\mu)\nu} = 0\, .
\ee

The dual of $\boldsymbol{k}$ is called
a principal tensor, that is
a rank two closed conformal Killing-Yano tensor $\boldsymbol{h}$
\be
h_{\mu\nu} = \frac{1}{2}e_{\mu\nu\alpha\beta}k^{\alpha\beta}\,,
\ee
where $e_{\mu\nu\alpha\beta}$ is the Levi-Civita tensor, with $e_{\tau\rho\theta\phi}=\sqrt{-g}$.
The tensor $h_{\mu\nu}$ is antisymmetric and obeys the equation\footnote{
The existence of a closed conformal Killing-Yano tensor is a generic property not only of the four-dimensional Kerr metric, but also of solutions of higher dimensional Einstein equations describing higher dimensional rotating black holes.
This tensor is a generator of the hidden symmetries, and it is often called the principal tensor.
Using the principal tensor, one can construct a tower of Killing tensors, which guarantee that the geodesic equations in such a spacetime are completely integrable. A discussion of this subject and additional references can be found in \cite{FRO_ZEL,LIVING}.}
\be \n{PT}
\nabla_{\lambda}h_{\mu\nu} = g_{\lambda\mu}\xi_{\nu} - g_{\lambda\nu}\xi_{\mu}\,,
\ee
where $\xi^{\mu}=\frac{1}{3}h^{\nu \mu}_{\ \ ;\nu}$.
One can check that $d\ts{h}=0$, and that $\ts{\xi}$ coincides with the Killing vector $\pa_\tau$.

Since $\boldsymbol{h}$  is a closed two-form, it can always be locally described by some potential $\boldsymbol{b}$
\be
h_{\mu\nu} = \partial_{\mu}b_{\nu}-\partial_{\nu}b_{\mu}\,.
\ee
In the Kerr metric in Boyer-Lindquist coordinates, the potential $\ts{b}$ is
\ba
b_{\mu} = -\frac{1}{2}\Big(&(\rho^2-\alpha^2\cos^2\theta)\delta_{\mu}^{\tau}\\
&-\alpha(\rho^2\sin^2\theta-\alpha^2\cos^2\theta)\delta_{\mu}^{\phi}\Big)\,.
\ea
The corresponding rank two closed conformal Killing-Yano tensor is
\ba
h_{\mu\nu} = &\rho(\delta_{\mu}^{\tau}\delta_{\nu}^{\rho}-\delta_{\mu}^{\rho}\delta_{\nu}^{\tau}) +
\alpha^2\sin\theta\cos\theta(\delta_{\mu}^{\tau}\delta_{\nu}^{\theta}-\delta_{\mu}^{\theta}\delta_{\nu}^{\tau}) \\
&+ \alpha\sin\theta\cos\theta(\rho^2+\alpha^2)(\delta_{\mu}^{\theta}\delta_{\nu}^{\phi}-\delta_{\mu}^{\phi}\delta_{\nu}^{\theta})\\
&+ \alpha\rho\sin^2\theta(\delta_{\mu}^{\rho}\delta_{\nu}^{\phi}-\delta_{\mu}^{\phi}\delta_{\nu}^{\rho})\,.
\ea

Using the inverse Hodge duality transformation, one gets
\be
k_{\mu\nu} = -\frac{1}{2}e_{\mu\nu\alpha\beta}h^{\alpha\beta}\,.
\ee
In the Kerr metric it has the following form
\ba
k_{\mu\nu} = &-\alpha\cos\theta(\delta_{\mu}^{\tau}\delta_{\nu}^{\rho}-\delta_{\mu}^{\rho}\delta_{\nu}^{\tau}) + \alpha\rho\sin\theta(\delta_{\mu}^{\tau}\delta_{\nu}^{\theta}-\delta_{\mu}^{\theta}\delta_{\nu}^{\tau}) \\
&- \alpha^2\sin^2\theta\cos\theta(\delta_{\mu}^{\rho}\delta_{\nu}^{\phi}-\delta_{\mu}^{\phi}\delta_{\nu}^{\rho})\\
&+ \rho\sin\theta(\rho^2+\alpha^2)(\delta_{\mu}^{\theta}\delta_{\nu}^{\phi}-\delta_{\mu}^{\phi}\delta_{\nu}^{\theta})\,.
\ea
Using \eqref{KKK} one finds
\be
K^{\mu}_{\hspace{5pt}\nu}=  \begin{bmatrix}
                    -\alpha^2 & 0 & 0 & \alpha\sin^2\theta(\rho^2+\alpha^2)\\
                    0 & -\alpha^2\cos^2\theta & 0 & 0\\
                    0 & 0 & \rho^2 & 0\\
                    -\alpha & 0 & 0 & \rho^2+\alpha^2\sin^2\theta
                \end{bmatrix}\,.
\ee

In addition to this Killing tensor, we can also construct a conformal Killing tensor $\ts{H}$ from our principal tensor
\be
H_{\mu\nu}=h_{\mu}^{\ \lambda}h_{\nu\lambda}\, .
\ee
In the Kerr metric it has the form
\be
H^{\mu}_{\hspace{5pt}\nu}=  \begin{bmatrix}
                    -(\rho^2+\alpha^2\sin^2\theta) & 0 & 0 & \alpha\sin^2\theta(\rho^2+\alpha^2)\\
                    0 & -\rho^2 & 0 & 0\\
                    0 & 0 & \alpha^2\cos^2\theta & 0\\
                    -\alpha & 0 & 0 & \alpha^2
                \end{bmatrix}\,.
\ee

One has
\be
H_{\mu\nu}=K_{\mu\nu}-\dfrac{1}{2} g_{\mu\nu} K_{\alpha}^{\alpha}\, .
\ee

We also define the following two tensors,
which will be useful in what follows.
The first one is
\be\n{HHKK}
\tilde{H}^{\mu}_{\hspace{5pt}\nu} = h^{\mu\lambda}k_{\nu\lambda}=\alpha\rho\cos\theta \delta^{\mu}_{\nu}\,.
\ee
The second tensor is
\be
\CAL{H}^{\mu}_{\ \nu}={H}^{\mu}_{\ \alpha}{H}^{\alpha}_{\ \nu}\, .
\ee
In the coordinates $(\tau,\rho,\theta,\phi)$ it has the following components
\be
\CAL{H}^{\mu}_{\ \nu}=\begin{bmatrix}
                    \rho^4+P\alpha^2\sin^2\theta & 0 & 0 & -P\alpha\sin^2\theta(\rho^2+\alpha^2)\\
                    0 & \rho^4 & 0 & 0\\
                    0 & 0 & \alpha^4\cos^4\theta & 0\\
                    P\alpha & 0 & 0 & -\alpha^2(P-\rho^2\cos^2\theta)
                \end{bmatrix}\,,
\ee
where $P=\rho^2-\alpha^2\cos^2\theta$.

\section{Null geodesics and associated  null tetrads} \label{s3}

\subsection{Parallel transport} \label{s3a}

Our main goal is study the propagation of polarized light in the Kerr spacetime. In the geometric optics approximation, light rays are null geodesics. The worldline of such a ray is described by the equation $x^{\mu}=x^{\mu}(\lambda)$. For an affine parametrization, its null tangent vector
$l^{\mu}={dx^{\mu}}/{d\lambda}$
satisfies \eqref{geodl}. We always assume that $\ts{l}$ is future-directed. In what follows, we shall use a set of vectors
$(\ts{l},\ts{n},\ts{e}_2,\ts{e}_3)$ associated with the null ray. We require that the vectors of this associated tetrad are parallel propagated along the null ray.  They should also satisfy the following normalization conditions
\be \n{lnee}
l^{\mu}n_{\nu}=-1\hh {e}_2^{\mu}{e}_{2 \mu}={e}_3^{\mu}{e}_{3 \mu}=1\, ,
\ee
where all other scalar products vanish. The first of the relations in \eqref{lnee} implies that the null vector $\ts{n}$ is also future-directed.

To construct the vectors
$(\ts{l},\ts{n},\ts{e}_2,\ts{e}_3)$, we follow the prescription described in \cite{Connell:2008vn}\footnote{Let us note that
a similar method allows one to obtain a set of parallel propagated vectors along timelike geodesics. The parallel propagated vectors for null and timelike geodesics were first constructed by Marck \cite{MARCK_1,MARCK_2}. For the generalization to the case of higher-dimensional rotating black holes, see \cite{Connell:2008vn,Kubiznak:2008zs}. Additional information on this subject can be found in \cite{FRO_ZEL,LIVING}.
}. We start by defining the following three vectors
\be \n{TILD}
\begin{split}
&e_2^{\mu}=\dfrac{1}{\sqrt{Q}} k^{\mu \nu} l_{\nu}\, ,\\  &\tilde{e}_{3}^{\mu}=\dfrac{1}{\sqrt{Q}} h^{\mu \nu} l_{\nu}\, ,\\
&\tilde{n}^{\mu}=\dfrac{1}{\sqrt{Q}}\big(
h^{\mu \nu} \tilde{e}_{3\nu}+\gamma l^{\mu}\big)\, .
\end{split}
\ee
Using the relation
\be
H_{\mu\nu}l^{\mu}l^{\nu}=K_{\mu\nu}l^{\mu}l^{\nu}=Q\,,
\ee
one can show that $e_2^{\mu} e_{2\mu}=\tilde{e}_3^{\mu} \tilde{e}_{3\mu}=1$.
Using \eqref{HHKK} one also gets
$e_{2\mu}\tilde{e}_3^{\mu}=0$.
The vectors $e_2^{\mu}$ and  $\tilde{e}_3^{\mu}$ are evidently orthogonal to $\ts{l}$. One also has
\be
l_{\mu}\tilde{n}^{\mu}=-1\, .
\ee
It is evident that $\tilde{e}_3^{\mu}\tilde{n}_{\mu}=0$. One also has
\be
{e}_2^{\mu}\tilde{n}_{\mu}=\dfrac{1}{Q^{3/2}}\tilde{H}^{\mu}_{\ \lambda}h^{\lambda\nu}l_{\mu}l_{\nu}=0\, .
\ee
The last equality follows from \eqref{HHKK}. One can check that the vector $\tilde{\ts{n}}$ is null provided
\be
\gamma=\frac{1}{2Q^{3/2}}\mathcal{H}_{\mu\nu}l^{\mu}l^{\nu}=-\frac{P}{2\sqrt{Q}}\, .
\ee
The constructed tetrad $(\ts{l},\tilde{\ts{n}},\ts{e}_2,\tilde{\ts{e}}_3)$ possesses the required normalization conditions \eqref{lnee}, but only two of its vectors, $\ts{l}$ and $\ts{e}_2$, are parallel propagated along the null ray $\ts{l}$.
This property of $\ts{e}_2$ can be demonstrated as follows. Let us denote
\be
D=l^{\alpha}\nabla_{\alpha}\, .
\ee
Then
\be \n{ee22}
D e_{2\mu} =
\frac{1}{\sqrt{Q}}\big(l^{\alpha}l^{\nu} k_{\mu\nu ;\alpha}
+l^{\alpha}l^{\nu}_{\ ;\alpha} k_{\mu\nu}\big)\, .
\ee
Equation
\eqref{kkk} implies that the first term in the right-hand side vanishes. The second term in the right-hand side also vanishes since $l^{\mu}$ is a tangent vector to a null geodesic.

Using property \eqref{PT} of the principal tensor
\be
Dh_{\mu\nu}=l_{\mu}\xi_{\nu}-l_{\nu}\xi_{\mu}\,,
\ee
one can show that
\be
D \tilde{e}_{3\mu}= \frac{1}{\sqrt{Q}}(\ts{\xi}\cdot \ts{l}) l_{\mu} = -\frac{1}{\sqrt{Q}}l_{\mu}\, ,
\ee
where in the last equality equation \eqref{EL1} was used. Therefore, the vector $\tilde{\ts{e}}_{3}$ is not parallel propagated. However, it is easy to ``upgrade" it to get a parallel propagated vector $\ts{e}_3$. Namely, let us  denote
\be \n{ee33}
{e}_{3}^{\mu}=\tilde{e}_{3}^{\mu}+\Phi l^{\mu}\hh
\Phi=\dfrac{1}{\sqrt{Q}}(\lambda+\lambda_0)\, ,
\ee
where $\lambda$ is the affine parameter along the null geodesic ray, and $\lambda_0$ is a constant.
Then it is easy to check that the vector ${e}_{3}^{\mu}$ is parallel propagated.

Similarly, one can show that the vector
\be \n{nnPP}
n^{\mu}=\tilde{n}^{\mu}+\Phi \tilde{e}_3^{\mu}+\dfrac{1}{2}\Phi^2 l^{\mu}
\ee
is parallel propagated along the null ray.
One can also check that the vectors of the tetrad $(\ts{l},\ts{n},\ts{e}_2,\ts{e}_3)$ satisfy the normalization conditions \eqref{lnee}\footnote{
Marck  constructed a similar tetrad in the Kerr-Newman spacetime in \cite{MARCK_1}. He first found the vector $\ts{e}_2$ by utilizing the Killing-Yano tensor. He found the two other vectors of the tetrad by ``educated guesses". The method proposed in \cite{Connell:2008vn} was used by Dahal \cite{Dahal:2023ncl} to construct the parallel propagated tetrad in the Kerr metric. It should be emphasized that the vectors of his tetrad $(\ts{l},\ts{n}, \ts{u}, \ts{v})$ in his notation differ from the vectors of our tetrad by the normalization they chose. The vector $\ts{l}$ has dimensions of $[Length^{-1}]$, while the vectors $\ts{u}$ and $\ts{v}$ have dimensions of $[Length^2]$ and the vector $\ts{n}$ has the dimension of $[Length^3]$. As a result, their normalization conditions differ from those given in \eqref{lnee}.
}.

\subsection{Geodesic null rays in the Kerr spacetime} \label{s3b}

Using relations \eqref{EL1} and \eqref{QQ}, one can find components of the vector $l^{\mu}=(\dot{\tau},\dot{\rho},\dot{\theta},\dot{\phi})$ tangent to the null geodesic $x^{\mu}(\lambda)$ in terms of its integrals of motion
\ba\n{llll}
&\Sigma\dot{\tau} = -\alpha\big(\alpha\sin^2\theta-\ell_{z}\big)+\frac{\rho^2+\alpha^2}{\Delta}\big(\rho^2+\alpha^2-\ell_{z}\alpha\big)\, , \\
&\Sigma\dot{\rho} = \epsilon_\rho \sqrt{\mathcal{R}}\,,\\
&\Sigma\dot{\theta} = \epsilon_{\theta}\sqrt{\Theta}\,,\\
&\Sigma\dot{\phi} = -\Big(\alpha-\frac{\ell_z}{\sin^2\theta}\Big)+\frac{\alpha}{\Delta}\big(\rho^2+\alpha^2-\ell_{z}\alpha\big)\, .
\ea
Here
\ba \n{RT}
\mathcal{R} = &\Big[\rho^2+\alpha^2-\alpha\ell_z\Big]^2-Q\Delta\,,\\
\Theta = &Q-\Big(\alpha\sin\theta-\frac{\ell_z}{\sin\theta}\Big)^2\,.
\ea
A dot over a variable signifies its derivative with respect to the affine parameter $\lambda$.
Since $\sqrt{\Theta}$ enters in the equations of motion, $\Theta$ should be non-negative, so that $Q\ge 0$.
This set of equations determines a congruence of null geodesics with fixed parameters $\ell_z$ and $Q$.
For completeness, we have also included the covariant form of the vectors contained in this tetrad in Appendix~\ref{apxA}.
The affine parameter $\lambda$ as a function of $\rho$ and $\theta$ is defined as follows
\be \n{LLL1}
\lambda =\epsilon_\rho \int^{\rho}\frac{\epsilon_\rho\rho^2d\rho}{\sqrt{\mathcal{R}}} +\int^{\theta}\frac{\epsilon_{\theta}\alpha^2\cos^2\theta d\theta}{\sqrt{\Theta}}\,.
\ee

In what follows, we focus on null geodesics which describe a photon being scattered by the black hole. Any such geodesic has one radial turning point at $\CAL{R}=0$. The sign parameter $\epsilon_\rho$ takes the value $\epsilon_\rho=-1$ for the incoming branch of the null ray before it reaches the turning point. After this point, for the outgoing branch $\epsilon_\rho=+1$.
Similarly, $\epsilon_{\theta}=-1$ for motion with a decrease in $\theta$, and $\epsilon_{\theta}=1$ for the part of the trajectory where $\theta$ increases.
Null geodesics in the Kerr geometry and their properties have been discussed in many publications\footnote{See e.g. \cite{Carter:68,Bardeen,MTW,Chandrasekhar}. An early review of the theory of geodesics in black hole spacetimes and
a rather complete set of the corresponding references can be found in \cite{Sharp}. Additional information and references to more recent publications can be found, e.g., in \cite{Cieslik}.}.

The parameters $\ell_z$ and $Q$ are directly related to the impact parameters of the photon trajectory, which we denote by $\alpha_{\theta}$ and $\alpha_{\phi}$. Namely, consider a ZAMO located at a far distance $\rho_0$ from the black hole along an angle $\theta_0$. Assume that this observer measures the directions of the photons relative to the center of symmetry of the asymptotic spacetime.
Let us denote the following components of $\ts{l}$ in the ZAMO frame \eqref{EEEE}
\be
l_{(\theta)}=E_{(\theta)}^{\mu}l_{\mu}\, \  \mbox{\ and\ } l_{(\phi)}=E_{(\phi)}^{\mu}l_{\mu}\,.
\ee
Then it is possible to show that in the limit $\rho_0\to \infty$, the two impact parameters have the following limits
\be
\begin{split} \n{imParam}
\alpha_{\theta}&=\lim_{\rho_0\to\infty}\rho_0 l_{(\theta)}=\pm\Big(Q-\big(\alpha\sin\theta-\ell_z/\sin\theta \big)^2\Big)^{1/2}\,,\\
\alpha_{\phi}&=\lim_{\rho_0\to\infty}-\rho_0 l_{(\phi)}=-\dfrac{\ell_z}{\sin\theta_0}\, .
\end{split}
\ee

A photon with the parameters $(\alpha_{\theta},\alpha_{\phi})$ is represented by a point in the image plane \cite{Bardeen}.
In particular, the photons with impact parameters passing at the wedge of the black hole shadow surface are represented by an image of the black hole's shadow.

\subsection{Null tetrad associated with a null geodesic ray} \label{s3c}

Using the expressions for
the vectors ${e}_2^{\mu}$, $\tilde{e}_3^{\mu}$ and $\tilde{n}^{\mu}$ given in \eqref{TILD}, one find the following explicit expressions for these vectors in the Kerr spacetime
\ba\n{eeee}
e_2^{\mu} = &\frac{1}{\Sigma\Delta\sqrt{Q}}\bigg(\alpha\cos\theta(\rho^2+\alpha^2)\epsilon_{\rho}\sqrt{{\mathcal{R}}}-\alpha\rho\Delta\sin\theta\epsilon_{\theta}\sqrt{{\Theta}},\\
&\hspace{42pt}\alpha\Delta\cos\theta\big((\rho^2+\alpha^2)-\alpha\ell_z\big),\\
&\hspace{42pt}-\rho\Delta\sin\theta\big(\alpha-\frac{\ell_z}{\sin^2\theta}\big),\\
&\hspace{42pt}\alpha^2\cos\theta\epsilon_{\rho}\sqrt{{\mathcal{R}}}-\rho\Delta\csc\theta\epsilon_{\theta}\sqrt{{\Theta}}\bigg)\, ,\\
\tilde{e}_{3}^{\mu} = &\frac{-1}{\Sigma\Delta\sqrt{Q}}\bigg(\rho(\rho^2+\alpha^2)\epsilon_{\rho}\sqrt{{\mathcal{R}}}+\alpha^2\sin\theta\cos\theta\Delta\epsilon_{\theta}\sqrt{{\Theta}},\\
&\hspace{42pt}\rho\Delta\big((\rho^2+\alpha^2)-\alpha\ell_z\big),\\
&\hspace{42pt}\alpha\Delta\sin\theta\cos\theta\big(\alpha-\frac{\ell_z}{\sin^2\theta}\big),\\
&\hspace{42pt}\alpha\rho\epsilon_{\rho}\sqrt{{\mathcal{R}}}+\alpha\Delta\cot\theta\epsilon_{\theta}\sqrt{{\Theta}}\bigg)\,, \\
\tilde{n}^{\mu}=&\frac{1}{2Q}\bigg(
\alpha(\alpha\sin^2\theta-\ell_z)+\frac{\rho^2+\alpha^2}{\Delta}(\rho^2+\alpha^2-\ell_{z}\alpha),\\
&\hspace{42pt} \epsilon_\rho \CAL{R}, -\epsilon_{\theta}\Theta,\\
&\hspace{42pt} \big(\alpha-\frac{\ell_z}{\sin^2\theta}\big)+\frac{\alpha}{\Delta}(\rho^2+\alpha^2-\ell_{z}\alpha)\bigg)\, .
\ea
The parallel propagated versions of vectors $\ts{\tilde{e}}_{3}$ and $\ts{\tilde{n}}$ are obtained by using equations \eqref{ee33} and \eqref{nnPP}. Let us note that we will mainly use the parallel propagated null tetrad $(\ts{l},\ts{n},\ts{e}_2,\ts{e}_3)$ in what follows. It is easy to write the components of this tetrad in an explicit form by using the relations \eqref{ee33}, \eqref{nnPP} and \eqref{eeee}. Since the corresponding expressions for $\ts{n}$ and $\ts{e}_3$ are rather long, we do not reproduce them in the paper. Let us also note that in many of the calculations that follow, it is sufficient and more convenient to use the representations \eqref{ee33} and \eqref{nnPP} for the vectors $\ts{n}$ and $\ts{e}_3$.

One can check that the constructed tetrad satisfies the relation
\be
e_{\mu\nu\alpha\beta}l^{\mu}n^{\nu}e_2^{\alpha}e_3^{\beta}=-1\, ,
\ee
and hence it has right-handed orientation.

Let us consider a null ray being scattered by the black hole. Its orbit has one radial turning point. We denote its coordinate by $(\rho_m,\theta_m)$, where $\theta_m$ is the value of $\theta$ at the radial turning point, not to be confused with its angular turning point where $\Theta=0$. To fix an ambiguity in the choice of the affine parameter $\lambda$, we put its value equal to zero for a radial turning point in the equatorial plane and write its the expression \eqref{LLL1} in the form
\be \n{LLL}
\begin{split}
&\lambda =\epsilon_\rho \Lambda(\rho)+F(\theta)\, ,\\
&\Lambda(\rho)=\int_{\rho_m}^{\rho}\dfrac{1}{\sqrt{\CAL{B}}}\, ,\\
&F(\theta)=\int_{\theta_m}^{\theta} \dfrac{\epsilon_{\theta} \alpha^2 \cos^2\theta d\theta}{\sqrt{\Theta}}\, ,\\
& \CAL{B}=\dfrac{\CAL{R}}{\rho^4}=\Big(1-\dfrac{\alpha(\ell_z-\alpha)}{\rho^2}\Big)^2 -\dfrac{Q}{\rho^4}\Delta\, .
\end{split}
\ee

The expression for $\Lambda(\rho)$ can be written in the following useful form
\be \n{Lambda}
\begin{split}
\Lambda&=\rho-\int_{\rho}^{\infty}\Big[
\dfrac{1}{\sqrt{\CAL{B}}}-1
\Big] d\rho +\lambda_m\, ,\\
\lambda_m&=J_m-\rho_m\hh
J_m=\int_{\rho_m}^{\infty}\Big[
\dfrac{1}{\sqrt{\CAL{B}}}-1\Big] d\rho\, .
\end{split}
\ee
At large $\rho$ one has
\be
\Lambda=\rho +\lambda_m +O(1/\rho)\, .
\ee

\section{Geodesic null rays in the equatorial plane} \label{s4}

\subsection{Null geodesics in the equatorial plane} \label{s4a}

In this section, we consider null geodesics and null tetrads associated with them in the equatorial plane of the Kerr metric.  For this case, many relations are greatly simplify. First, let us write down the dimensionless Kerr metric in the equatorial plane
\ba
ds^2 = &-\Big(1-\frac{2}{\rho}\Big)d\tau^2+\frac{\rho^2}{\Delta}d\rho^2-\frac{4\alpha}{\rho}d\tau d\phi\\
&+\rho^2 d\theta^2+\Big(\rho^2+\alpha^2+\frac{2\alpha^2}{\rho}\Big)d\phi^2\,.
\ea
It is easy to show that the black hole ergosphere crosses the equatorial plane at $\rho=2$.

We also have the following nonzero components of the Riemann tensor in the equatorial plane
\ba
R_{\tau\rho\tau\rho} = &-\frac{2\Delta+\alpha^2}{\Delta\rho^3}\hhh R_{\tau\theta\tau\theta}=\frac{\Delta+2\alpha^2}{\rho^3}\hhh  R_{\tau\phi\tau\phi}=\frac{\Delta}{\rho^3}\,,\\
R_{\tau\rho\rho\phi}=&-\frac{\alpha(3\Delta+2\rho)}{\Delta\rho^3}\hhh R_{\tau\theta\theta\phi}=\frac{\alpha(3\Delta+4\rho)}{\rho^3}\,,\\
R_{\rho\theta\rho\theta}=&-\frac{\rho}{\Delta}\hhh R_{\rho\phi\rho\phi}=-\frac{(\rho^2+\alpha^2)^2+2\Delta\alpha^2}{\Delta\rho^3}\,,\\
R_{\theta\phi\theta\phi}=&\frac{2(\rho^2+\alpha^2)^2+\Delta\alpha^2}{\rho^3}\,.
\ea

Let us now look at the null geodesics\footnote{
Useful information about null geodesics in the Kerr spacetime can be found e.g. in \cite{FRO_ZEL}.
}. For null geodesics in the equatorial plane, $\Theta=0$, implying that
\be \n{QQQQ}
Q=(\ell_z-\alpha)^2\, .
\ee
This implies that the ``Carter constant" $C$ defined as $C=Q-(\ell_z-\alpha)^2$ vanishes for null geodesics on the equatorial plane, as it should.

Let us denote
\be\n{BBDD}
\CAL{B}=1-\dfrac{\ell_z^2-\alpha^2}{\rho^2}+\dfrac{2(\ell_z-\alpha)^2}{\rho^3}\, .
\ee
Then the tangent vector to a null geodesic in the equatorial plane can be written as follows
\be  \n{tetradEQ}
\begin{split}
& l^{\mu}=(\dot{\tau},\dot{\rho},0,\dot{\phi})\, ,\\
&\dot{\tau}=\dfrac{\rho(\rho^2+\alpha^2)-2\alpha(\ell_z-\alpha)}{\rho \Delta}\, ,\\
&\dot{\rho}=\epsilon_{\rho} \sqrt{\CAL{B}}\, ,\\
&\dot{\phi}=\dfrac{
(\rho-2)\ell_z +2\alpha
}{\rho \Delta}\, .
\end{split}
\ee

Every null geodesic has no more than one turning point. Such a turning point $\rho=\rho_m$, where $\CAL{B}=0$, exists for null geodesics that come from infinity and then bounce back to infinity.
One can solve the equation $\CAL{B}=0$ and get
\be \n{LLRR}
\ell^{\pm}_z=\dfrac{\pm \rho_m\sqrt{\Delta_m}-2\alpha}{\rho_m-2}\hh \Delta_m=\rho_m^2-2\rho_m+\alpha^2
\, .
\ee
We choose the direction of the rotation of the black hole so that the dimensionless rotation parameter $\alpha$ is non-negative, $0\le \alpha \le 1$.
It is possible to check that for sign +
in \eqref{LLRR}
the dimensionless impact parameter $\ell^+_z$ is positive. This case corresponds to {\it{prograde}} motion of the null ray. For sign $-$, $\ell^-_z$ is negative and exhibits {\it{retrograde}} motion.
For null rays which travel far away from the black hole one has
\be \n{rmlarge}
\ell_z^{\pm}\approx \pm(\rho_m +1)\, .
\ee

\begin{figure}[!hbt]
    \centering
\includegraphics[width=0.45\textwidth]{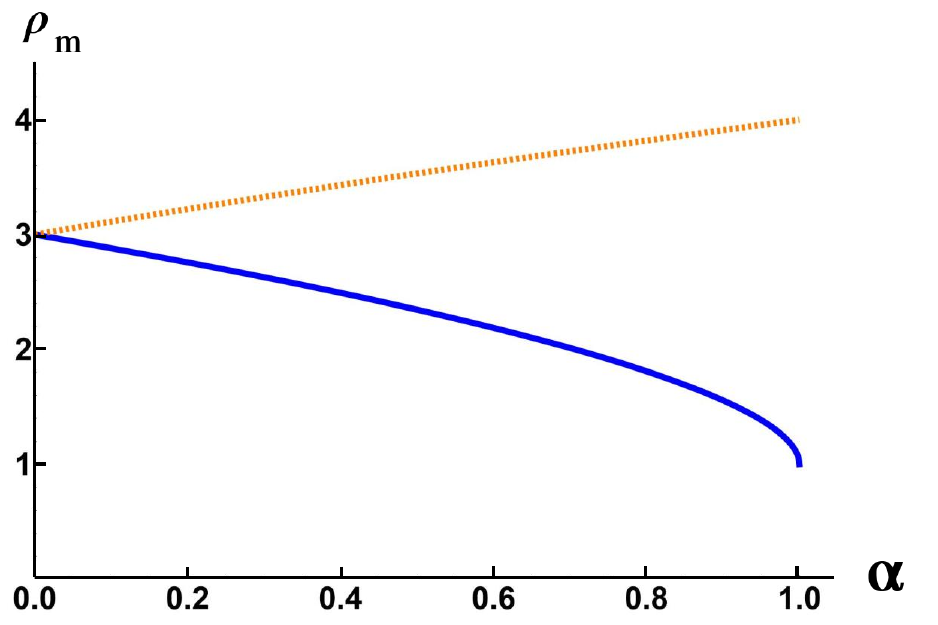}
\vspace{0.25cm}
\caption{\n{rm_a}A plot of $\rho^{\pm}_m$ as a function of the dimensionless rotation parameter $\alpha$. The lower branch of the curve (solid) gives $\rho^{+}_m$ for prograde null rays, while the upper branch of the curve (dashed) gives $\rho^{-}_m$ for retrograde null rays.}
\end{figure}

Simple analysis shows that, for $\rho_m>2$, $\ell^+_z$ as a function of $\rho_m$ has a minimum at some critical value $\rho_m=\rho_m^+$, while $\ell_z^-$ has a maximum at $\rho_m=\rho_m^-$, where \cite{FRO_ZEL}
\be \n{rhopm}
\rho_m^{\pm}=2+\cos\chi \mp \sqrt{3}\sin\chi\hh \chi =\dfrac{2}{3}\arcsin\alpha\, .
\ee
This relation is displayed graphically in Fig.~\ref{rm_a}.

\subsection{Associated null tetrad}

On the equatorial plane, the vectors $\ts{e}_2$, $\tilde{\ts{e}}_3$ and $\tilde{\ts{n}}$ defined by \eqref{eeee} take the form
\be\n{TETR}
\begin{split}
e_2^{\mu}\partial_{\mu} =& (1/\rho)\partial_{\theta}\,,\\
\tilde{e}_3^{\mu}\partial_{\mu}=&-
\dfrac{\epsilon_\rho \rho(\rho^2+\alpha^2)\sqrt{\CAL{B}}}{(\ell_z-\alpha)\Delta}\partial_{\tau}-\dfrac{\rho^2-\alpha(\ell_z-\alpha)}{(\ell_z-\alpha)\rho}\partial_{\rho}\\
&-\dfrac{\epsilon_\rho \alpha \rho\sqrt{\CAL{B}}}{(\ell_z-\alpha)\Delta}\partial_{\phi}\, ,\\
\tilde{n}^{\mu}\partial_{\mu} = &\frac{1}{2(\ell_z-\alpha)^2}\bigg(\Big(\frac{(\rho^2+\alpha^2)(\rho^2-\alpha(\ell_z-\alpha))}{\Delta}\\
&-\alpha(\ell_z-\alpha)\Big)\partial_{\tau}+\epsilon_{\rho}\Big(\rho^2\sqrt{\CAL B}\Big)\partial_{\rho}\\
&+\Big(\frac{\alpha(\rho^2-\alpha(\ell_z-\alpha))}{\Delta}-(\ell_z-\alpha)\Big)\partial_{\phi}\bigg)\,.
\end{split}
\ee
The relations \eqref{TET_ZAM} in appendix~A give the expressions of these vectors in the ZAMO frame.

\begin{figure}[!hbt]
    \centering
\includegraphics[width=0.45\textwidth]{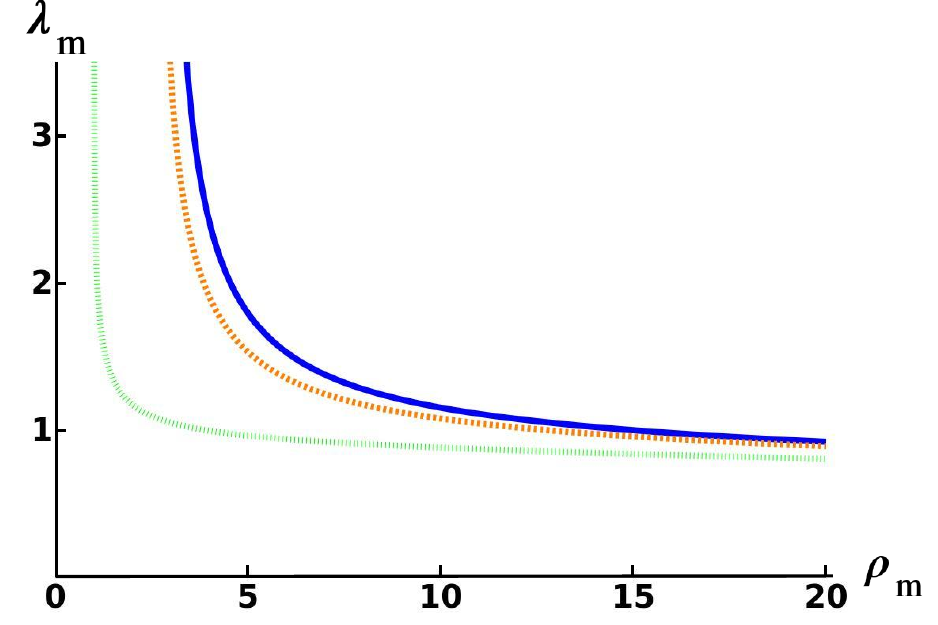}
\caption{\n{Lm_p}A plot of $\lambda_m=J_m-\rho_m$ as a function of $\rho_m$ for prograde photon scattering trajectories in the equatorial plane, where $\alpha =$ 0 (solid), 0.5 (dashed), 1 (dotted) enumerates each line going from right to left.}
\end{figure}

\begin{figure}[!hbt]
    \centering
\includegraphics[width=0.45\textwidth]{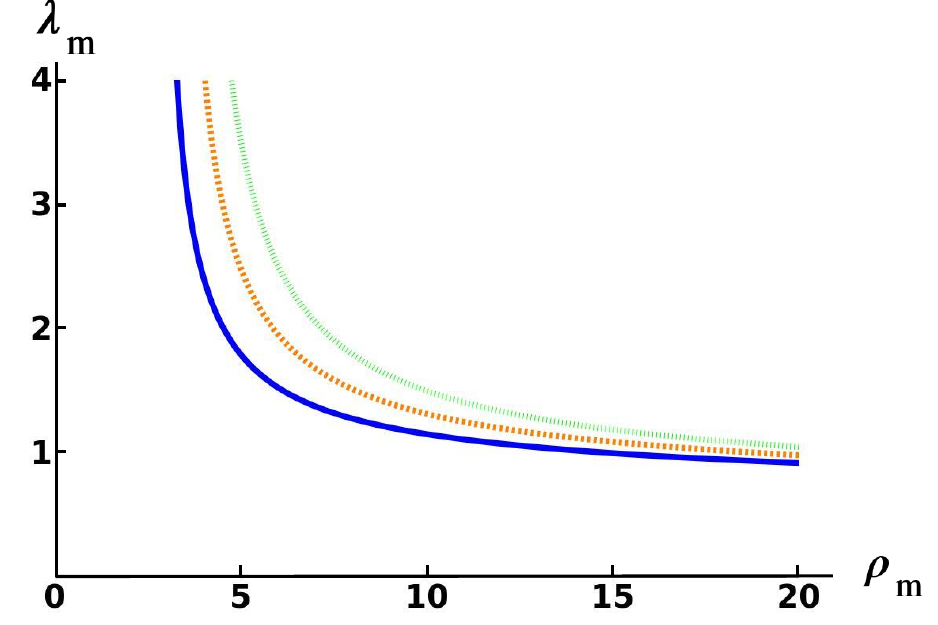}
\caption{\n{Lm_m}A plot of $\lambda_m=J_m-\rho_m$ as a function of $\rho_m$ for retrograde photon scattering trajectories in the equatorial plane, where $\alpha =$ 0 (solid), 0.5 (dashed), 1 (dotted) enumerates each line going from left to right.}
\end{figure}

In the equatorial plane, the function $F(\theta)$ equation \eqref{LLL} for the affine parameter $\lambda$ vanishes, and one has
\be \n{LambdaEQ}
\begin{split}
\Lambda&=\rho-\int_{\rho}^{\infty}\Big[
\dfrac{1}{\sqrt{\CAL{B}}}-1
\Big] d\rho +\lambda_m\, ,\\
\lambda_m&=J_m-\rho_m\hh
J_m=\int_{\rho_m}^{\infty}\Big[
\dfrac{1}{\sqrt{\CAL{B}}}-1\Big] d\rho\, .
\end{split}
\ee
One can see how the parameter $\lambda_m$ behaves for prograde and retrograde rays in Figs.~\ref{Lm_p} and \ref{Lm_m}, respectively. For large $\rho$, one has
\be
\Lambda=\rho+\lambda_m+O(1/\rho)\, .
\ee

For rays propagating far from the black hole where $\rho_m\gg 1$, one can use the approximate value of the impact parameter given in \eqref{BBDD}. Then
\be \n{Brm}
\begin{split}
&\CAL{B}\approx \dfrac{
\hat{\rho}^2-1}{\hat{\rho}^2}\big[
1-\dfrac{2}{\hat{\rho}(\hat{\rho}+1)}\dfrac{1}{\rho_m}
\big]\, ,\\
& \dfrac{1}{\sqrt{\CAL{B}}}-1\approx
\big(\dfrac{\hat{\rho}}{\sqrt{\hat{\rho}^2-1}}-1\big)\\
& \ \ \ +\dfrac{1}{(\hat{\rho}+1)^{3/2}(\hat{\rho}-1)^{1/2}}\dfrac{1}{\rho_m}\, .
\end{split}
\ee
Here we denote $\hat{\rho}=\rho/\rho_m$. Using \eqref{LambdaEQ}, one can write
\be
\lambda_m/\rho_m=\int_{1}^{\infty}\Big[
\dfrac{1}{\sqrt{\CAL{B}}}-1\Big] d\hat{\rho}-1\, .
\ee
The integral in this expression can easily be taken approximately. Calculations give
\be\n{LMR}
\lambda_m/\rho_m\approx 1/\rho_m\, .
\ee
Thus, for large $\rho_m$, the parameter $\lambda_m$ has the limit $\lambda_m=1$. The plots for $\lambda_m$ shown in Figs.~\ref{Lm_p} and \ref{Lm_m} are in agreement with this asymptotic property of $\lambda_m$.

The parallel propagated versions of the vectors $\ts{\tilde{n}}$ and $\ts{\tilde{e}}_3$ can be found by using \eqref{ee33} and \eqref{nnPP}.
We write the function $\Phi$ which enters these relations in the following form
\be\n{ZZZZ}
\Phi=\dfrac{\epsilon_\rho \Lambda(\rho)+\lambda_0}{\ell_z-\alpha}
\, .
\ee
The parameter $\lambda_0$ is a constant that should be defined by the initial conditions of the corresponding problem.

Using  expressions \eqref{TET_ZAM} for the decomposition of the tetrad $(\ts{l},\ts{n},\ts{e}_2,\ts{e}_3)$  in the ZAMO basis one can find the following relations valid in the  asymptotic domain $\rho\to\infty$
\ba\n{ASSIM}
l^{\mu}=&\big(E_{(\tau)}^{\mu}+\epsilon_{\rho}E_{(\rho)}^{\mu}\big)\big(1+1/\rho\big)+\ell_{z}E_{(\phi)}+...\,,\\
n^{\mu}=&\frac{1}{2}\big(E_{(\tau)}^{\mu}-\epsilon_{\rho}E_{(\rho)}^{\mu}\big)
+\dfrac{1}{2}\Delta\lambda^2 l^{\mu}\\
&+\epsilon_{\rho}\Delta\lambda E_{(\phi)}^{\mu}+ O(1/\rho)\,,\\
e_{2}^{\mu}=&E_{(\theta)}^{\mu}\,,\\
e_{3}^{\mu}=&\epsilon_{\rho}E_{(\phi)}^{\mu}+\Delta\lambda l^{\mu}+O(1/\rho)\,.
\ea
Here,
\be
\Delta\lambda=\frac{\lambda_0+\epsilon_{\rho}\lambda_m}{\ell_z-\alpha}\,.
\ee

We see that $\lambda_0=\lambda_m$ is a natural choice for scattering, as the null tetrad vectors $\ts{e}_2$ and $\ts{e}_3$ for incoming rays will ``coincide" with the ZAMO basis vectors $\ts{E}_{\theta}$ and $\ts{E}_{\phi}$ as $\rho\to\infty$.  This choice is adopted at some points later in the paper.

\section{Spinoptics in Kerr metric} \label{s5}

\subsection{Spinoptics equations} \label{s5a}

A ray describing photon motion in the geometric optics approximation is a null geodesic.
In the spinoptics approximation, due to the interaction of the photon's spin with the curvature of spacetime, its worldline is still null but no longer a geodesic. As a result, the parallel propagated null tetrad associated with these rays will slightly differ from the null tetrads associated with null geodesics. In order to distinguish these two different tetrads we need to change our notations. We use the subscript $0$ to indicate that the corresponding tetrad is associated with the null geodesics. Thus from now on, this tetrad, discussed in the previous sections will be denoted by $(\ts{l}_0,\ts{n}_0,\ts{e}_{0 2},\ts{e}_{0 3})$ , while  the other (non-geodesic) tetrad
is $(\ts{l},\ts{n},\ts{e}_{2},\ts{e}_{3})$. We shall also use the complex null vectors
\be\n{MM0}
m_0^{\mu}=\dfrac{1}{\sqrt{2}}(e_{2}^{0\mu}+ie_{03}^{\mu})\hh
\bar{m}_0^{\mu}=\dfrac{1}{\sqrt{2}}(e_{02}^{\mu}-ie_{03}^{\mu})\, .
\ee

The parallel transport equations have the form
\be \n{lnmm0}
D_0 l_0^{\mu}=D_0 n_0^{\mu}=D_0 m_0^{\mu}=
D_0 \bar{m}_0^{\mu}=0\, ,
\ee
where $D_0=l_0^{\alpha}\nabla_{\alpha}$.

In the spinoptics approximation  null rays obey the following equation \cite{Frolov:2020uhn}
\be \n{DDLLL}
\begin{split}
D \ell^{\mu}&=\chi \Upsilon^{\mu}\hh
D=\ell^{\alpha}\nabla_{\alpha}\, ,\\
\Upsilon^{\mu}&=iR_{\mu\nu\alpha\beta}\ell^{\nu}m^{\alpha}\bar{m}^{\beta}\hh
\chi=\dfrac{\sigma}{M\omega}
\,.
\end{split}
\ee
Here, as in previous papers, we define $\sigma$ to be the helicity factor. It takes the values $\pm 1$ for circularly-polarized electromagnetic waves and $\pm 2$ for circularly-polarized gravitational waves. The parameter $\omega$ is the frequency of the waves. In the adopted high-frequency approximation, the dimensionless coupling constant $\chi$ is assumed to be small.

The term $\chi \Upsilon^{\mu}$ on the right-hand side of the equation for a null ray is nothing but its four-dimensional acceleration induced by the interaction of the spin with the spacetime curvature. For $\chi=0$, it vanishes and the rays are geodesics.
Equation \eqref{DDLLL} has the same form as Equation \eqref{hatl} discussed in Appendix~\ref{apxB}. As it is demonstrated in this appendix, using the freedom of choice of a complex null tetrad, one can choose it so that the following equations are satisfied \eqref{EQL}
\be \n{EQLC}
\begin{split}
& Dl^{\mu}=\chi \Upsilon^{\mu}=w_0 l^{\mu}+
\bar{\kappa} m^{\mu}+{\kappa}\bar{m}^{\mu}\, ,\\
&Dn^{\mu} = w_0 n^{\mu}\hh Dm^{\mu}=\kappa n^{\mu}\,  .
\end{split}
\ee

In the high-frequency approximation, which is adopted in this paper, the parameter $\chi$ is small, and therefore the deviation of the corresponding null ray from the geodesic one is also small.
To find the corresponding approximate solution, we shall use the method of perturbations.
In this approximation, it is sufficient to use the complex null vectors that are parallel propagated along unperturbed null geodesics in the right-hand side of \eqref{DDLLL}. It is also convenient to use parallel propagated vectors $\ts{e}_{02}$ and $\ts{e}_{03}$ instead of $\ts{m}_0$ and $\bar{\ts{m}}_0$ (see \eqref{MM0}).
Under these assumptions, the ray equation \eqref{DDLLL} takes the form
\be \n{psi}
\begin{split}
Dl^{\mu}&=\chi (\psi_0 l_0^{\mu} +
\psi_e e_{02}^{\mu}+\psi_3 e_{03}^{\mu})\, ,
\end{split}
\ee
where
\be
\begin{split}
\psi_0&=-R_{\mu\nu\alpha\beta}n_0^{\mu}l_0^{\nu}e_{02}^{\alpha}e_{03}^{\beta}\, \\
\psi_2&=R_{\mu\nu\alpha\beta}e_{02}^{\mu}l_0^{\nu}e_{02}^{\alpha}e_{03}^{\beta}\, \\
\psi_3&=R_{\mu\nu\alpha\beta}e_{03}^{\mu}l_0^{\nu}e_{02}^{\alpha}e_{03}^{\beta}\, ,
\end{split}
\ee
where $R_{\mu\nu\alpha\beta}$ is the Riemann tensor for the Kerr metric. Calculations for $\psi_a$ give the following expressions
\be
\begin{split}
\psi_0&=\frac{\alpha\cos\theta}{\Sigma^5}\Big(\Sigma^2\big(3\rho^2-\alpha^2\cos^2\theta\big)\\
&-3Q\Phi^2\big(5\rho^4-10\rho^2\alpha^2\cos^2\theta+\alpha^4\cos^4\theta\big)\Big)\,, \\
\psi_2&=\frac{3\rho Q\Phi}{\Sigma^5}\big(\rho^4-10\rho^2\alpha^2\cos^2\theta+5\alpha^4\cos^4\theta\big) \,, \\
\psi_3&=\frac{3\alpha\cos\theta Q\Phi}{\Sigma^5}\big(5\rho^4-10\rho^2\alpha^2\cos^2\theta+\alpha^4\cos^4\theta\big)\, .
\end{split}
\ee

\section{Tilting effect near equatorial plane} \label{s6}

\subsection{Null ray deviation equation} \label{s5b}

In the presence of the ``driving force" $\chi \Upsilon^{\mu}$, the null rays propagate slightly differently than geodesic rays with $\chi=0$. To describe their deviation from geodesic rays, one can consider a null geodesic and suppose that at some point $O$ an ``accelerated ray" passing through this point has the same tangent vector $l^{\mu}$ as the null geodesic. Let us evolve the rays with the same canonical parameter $\lambda$, and denote the corresponding vector connecting these rays by $\delta x^{\mu}$. Then, for a small acceleration parameter $\chi$ in the leading order in $\delta x^{\mu}$, one has the following deviation equation
\be \n{0_DDLL}
D_0^2 \delta x^{\mu}-R^{\mu}_{\alpha\beta\gamma}l_0^{\alpha}l_0^{\beta}\delta x^{\gamma}=\chi \Upsilon^{\mu}\, .
\ee
Here $D_0=l_0^{\alpha}\nabla_{\alpha}$.
The derivation of this equation can be found in Appendix~\ref{DEV}. Let us note that in the absence of the driving force $\chi \Upsilon^{\mu}$, this equation correctly reproduces the standard equation for the deviation of null geodesics.

\subsection{Tilting effect for rotating black holes}

\begin{figure}[!hbt]
    \centering
\includegraphics[width=0.45\textwidth]{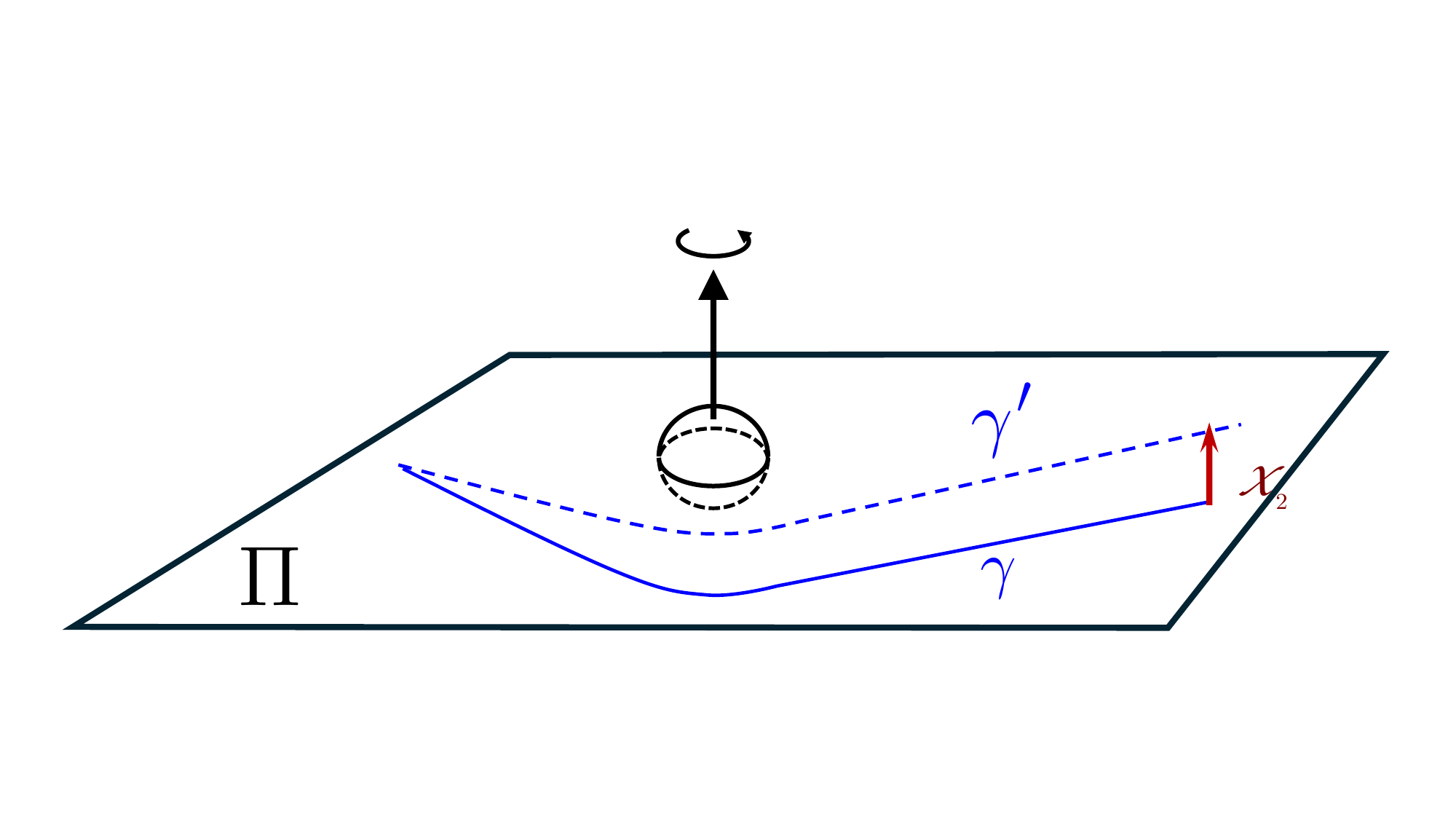}
%\vspace{-3cm}
\caption{\n{x2Fig}Displacement of the null ray near the equatorial plane.}
\end{figure}

We now apply the general equation \eqref{0_DDLL} for the deflection of null rays in the spinoptics approximation to the case where these rays propagate close to the equatorial plane. Namely, we demonstrate that the spinoptics effect results in the shift of the ray originally sent along the equatorial plane $\Pi$ away from it. This is schematically illustrated in Fig.~\ref{x2Fig}. The line $\gamma$ shows a geodesic null ray propagating along $\Pi$, while $\gamma'$ is a similar null ray but it follows the spinoptics approximation.

Let us denote
\be
\begin{split}
&\CAL{A}_{\mu\nu}=R_{\mu\alpha\beta\nu}l_0^{\alpha}l_0^{\beta}\, ,\\
& \Upsilon^{\mu}=\psi_0 l_0^{\mu} +
\psi_2 e_{02}^{\mu}+\psi_3 e_{03}^{\mu}\, .
\end{split}
\ee
Then equation \eqref{DDLL}, written in dimensionless form, reads
\be \n{EEDEF}
D_0^2 \delta x^{\mu} -\CAL{A}^{\mu}_{\ \nu}\delta x^{\nu}=\chi\Upsilon^{\mu} \, .
\ee

Calculations show that on the equatorial plane the only non-vanishing tetrad components of $\CAL{A}_{\mu\nu}$  are
\ba
\CAL{A}_{\mu\nu}n_0^{\mu}n_0^{\nu}&=\frac{1}{\rho^5}\big(3Q\Phi^2-\rho^2\big)\,,\\
\CAL{A}_{\mu\nu}e_{02}^{\mu}e_{02}^{\nu}&=-\CAL{A}_{\mu\nu}e_{03}^{\mu}e_{03}^{\nu}=-\frac{3Q}{\rho^5}\,.
\ea

For null rays propagating near the equatorial plane $\theta=\pi/2$, the angle $\vartheta=\pi/2-\theta$ is small.
The vector $e_{02}^{\mu}$ is directed along $\theta$, while $-e_{02}^{\mu}$ is directed along $\vartheta$, and hence it points towards the upper side of the equatorial plane. We denote by  $\CAL{X}_2$ the following displacement of the ray
\be \n{X2theta}
\CAL{X}_2=-\delta x^{\mu}e_{02\mu}=\rho \vartheta\, .
\ee

By contracting equation \eqref{EEDEF} with $e_{2\mu}$, one gets
\be \n{XX2}
\dfrac{d^2 \CAL{X}_2}{d\lambda^2}+\dfrac{3Q}{\rho^5} \CAL{X}_2=-\chi \psi_2\, .
\ee
In the equatorial plane, one has
\be \n{psi22}
\psi_2=\frac{3 Q\Phi}{\rho^5}\, .
\ee
\begin{figure}[!hbt]
    \centering
\includegraphics[width=0.45\textwidth]{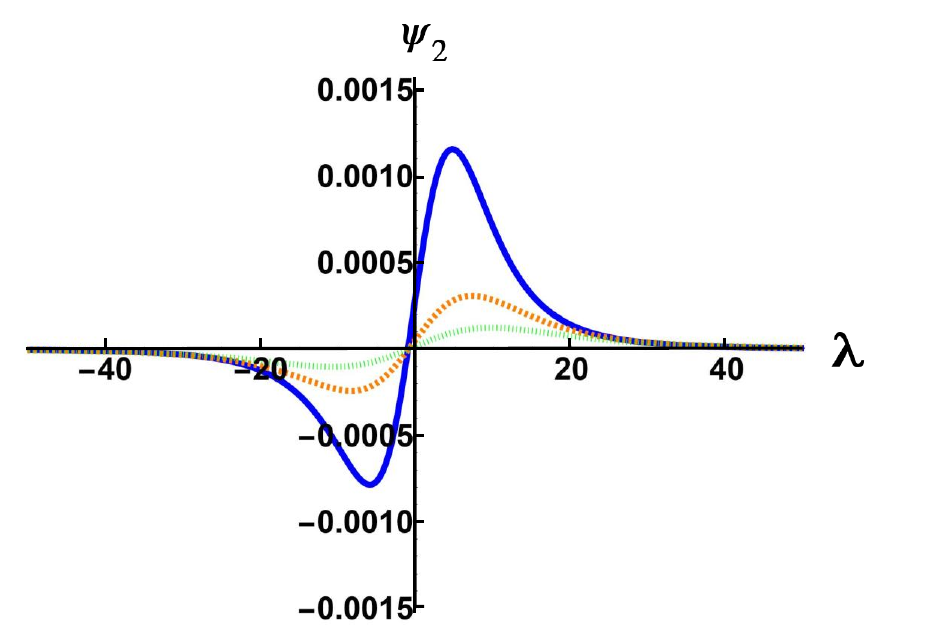}
\caption{\n{psi2p}Driving function $\psi_2$ for prograde rays as a function of $\lambda$ for $\alpha=0.5$ and $\rho_m$=10 (solid), 15 (dashed), and 20 (dotted).}
\end{figure}

\begin{figure}[!hbt]
    \centering
\includegraphics[width=0.45\textwidth]{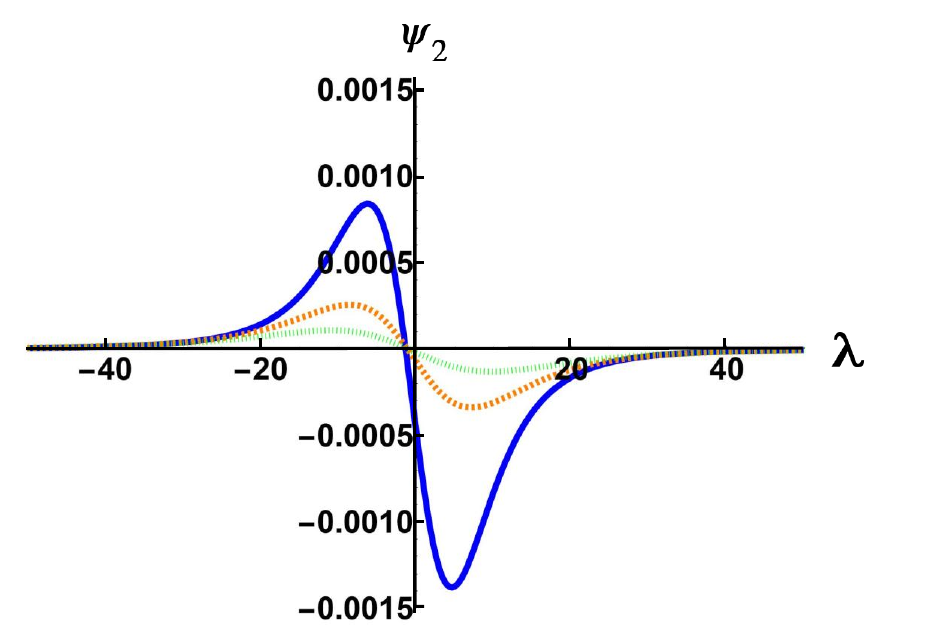}
\caption{\n{psi2m}Driving function $\psi_2$ for retrograde rays as a function of $\lambda$ for $\alpha=0.5$ and $\rho_m$=10 (solid), 15 (dashed), and 20 (dotted).}
\end{figure}
Relations \eqref{ASSIM} imply that for $\lambda_0=\lambda_m$, when $\Delta\lambda=0$, the polarization vector $\ts{m}_0$ for the incoming radiation has the standard form
\be
\ts{m}_0\approx \dfrac{1}{\sqrt{2}\rho}(\pa_{\theta} +i\pa_{\phi})\, .
\ee
We use this choice and write the function $\Phi$, which enters \eqref{psi22}, in the form
\be \n{PPPP}
\Phi=\dfrac{1}{\ell_z-\alpha}(\epsilon_\rho\Lambda(\rho)+\lambda_m)\, .
\ee

Using equations \eqref{psi22} and \eqref{PPPP} one gets
\be \n{PPSI}
\psi_2=\dfrac{3(\ell_z-\alpha)(\lambda+\lambda_m)}{\rho^5}\, .
\ee
Figs.~\ref{psi2p} and \ref{psi2m} show $\psi_2$ for $\alpha=0.5$ and several values of $\rho_m$ for both prograde and retrograde rays, respectively.

Let us note that if one puts $\lambda_m=0$ in \eqref{PPSI}, then the  $\psi_2=\psi_2(\lambda)$ would be an odd function of $\lambda$ and its plot would cross the $\lambda$ axis at $\lambda=0$. Due to the presence of the factor $\lambda_m$, the point where $\psi_2=0$ is slightly shifted to the point $\lambda=-\lambda_m$. To make this more visible, we present plots similar to \ref{psi2p} and \ref{psi2m} with magnified resolution of the region close to the origin, which are given by Figs.~\ref{psi2p2} and \ref{psi2m2}, respectively.

\begin{figure}[!hbt]
    \centering
\includegraphics[width=0.45\textwidth]{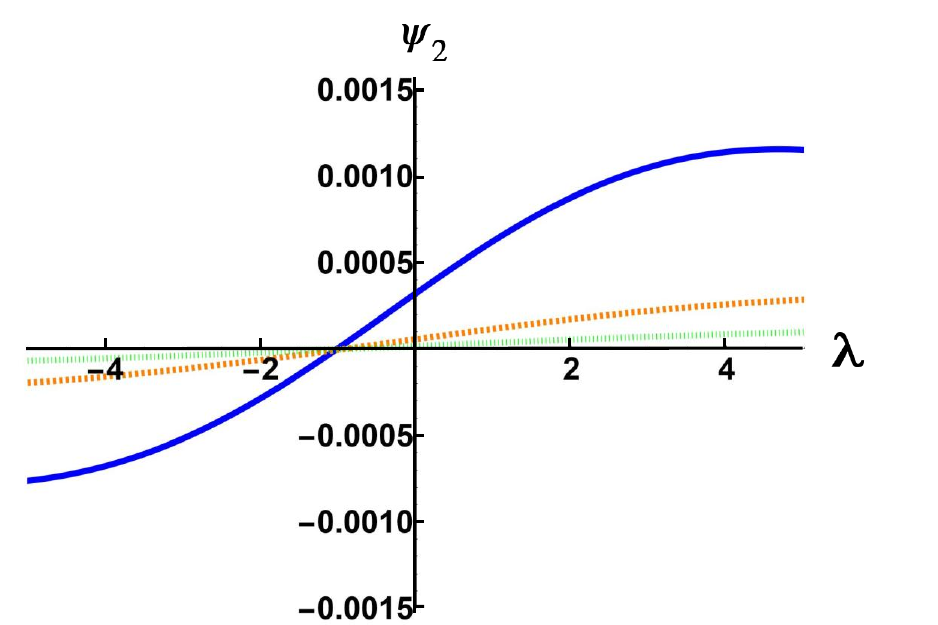}
\caption{\n{psi2p2}Close-up view of driving function $\psi_2$ for prograde rays as a function of $\lambda$ for $\alpha=0.5$ and $\rho_m$=10 (solid), 15 (dashed), and 20 (dotted).}
\end{figure}

\begin{figure}[!hbt]
    \centering
\includegraphics[width=0.45\textwidth]{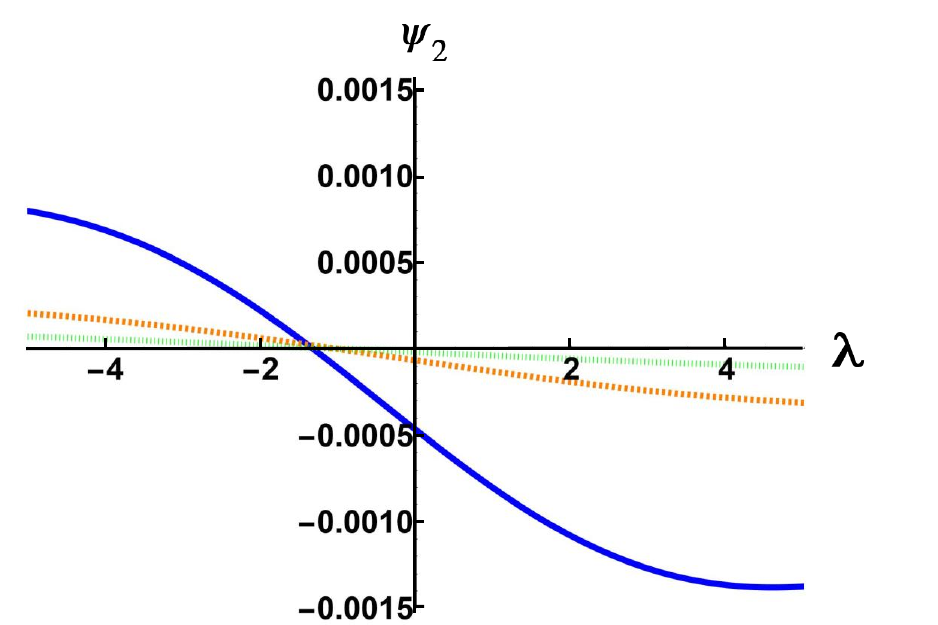}
\caption{\n{psi2m2}Close-up view of driving function $\psi_2$ for retrograde rays as a function of $\lambda$ for $\alpha=0.5$ and $\rho_m$=10 (solid), 15 (dashed), and 20 (dotted).}
\end{figure}

The equation \eqref{XX2} should be accompanied by the equation for $\rho=\rho(\lambda)$
\be \n{dotr}
\dot{\rho}=\mbox{sign}(\lambda) \sqrt{\CAL{B}}\,.
\ee
Since this equation is singular at the radial turning point $\rho=\rho_m$, we shall use the second order equation obtained from \eqref{dotr} by differentiating it with respect to $\lambda$
\be \n{ddotr}
\ddot{\rho}=
\dfrac{\ell_z-\alpha}{\rho^4}\Big[
(\rho-3)\ell_z+(\rho+3)\alpha\Big]\, .
\ee
Using \eqref{X2theta}, \eqref{PPSI}, \eqref{dotr}, and \eqref{ddotr} one obtains the following equation for the titling angle $\vartheta$
\be \n{ddotl}
\ddot{\vartheta}= -\frac{1}{\rho}\Big[2\dot{\rho}\dot{\vartheta}+\frac{\ell_z^2-\alpha^2}{\rho^3}\vartheta+\chi\psi_2\Big] \, .
\ee

\begin{figure}[!hbt]
    \centering
\includegraphics[width=0.45\textwidth]{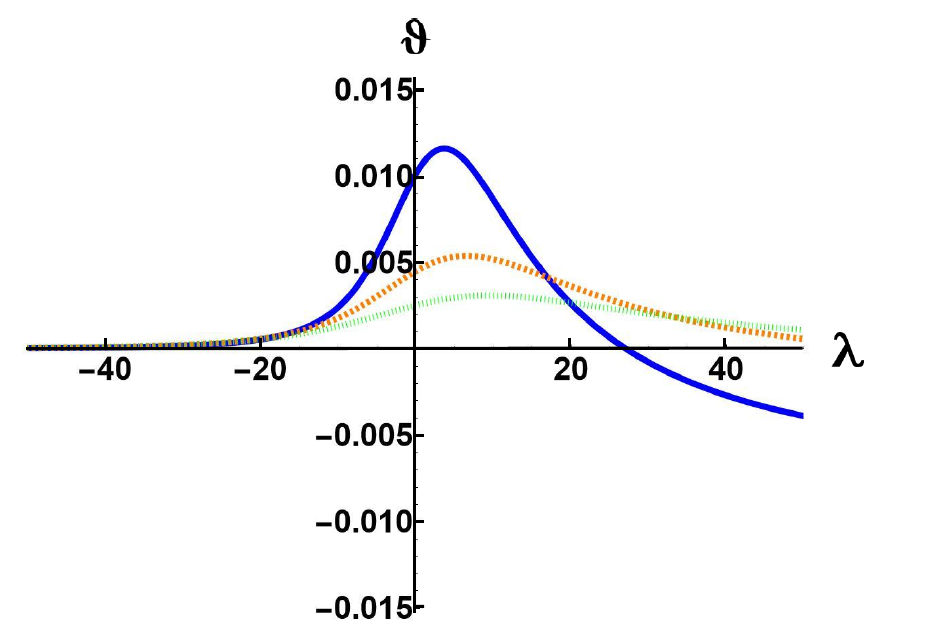}
\caption{\n{TP_l}A plot of $\vartheta(\lambda)$ for prograde rays with fixed rotation parameter $\alpha=0.5$ and varying $\rho_m$=10 (solid), 15 (dashed), and 20 (dotted).}
\end{figure}

\begin{figure}[!hbt]
    \centering
\includegraphics[width=0.45\textwidth]{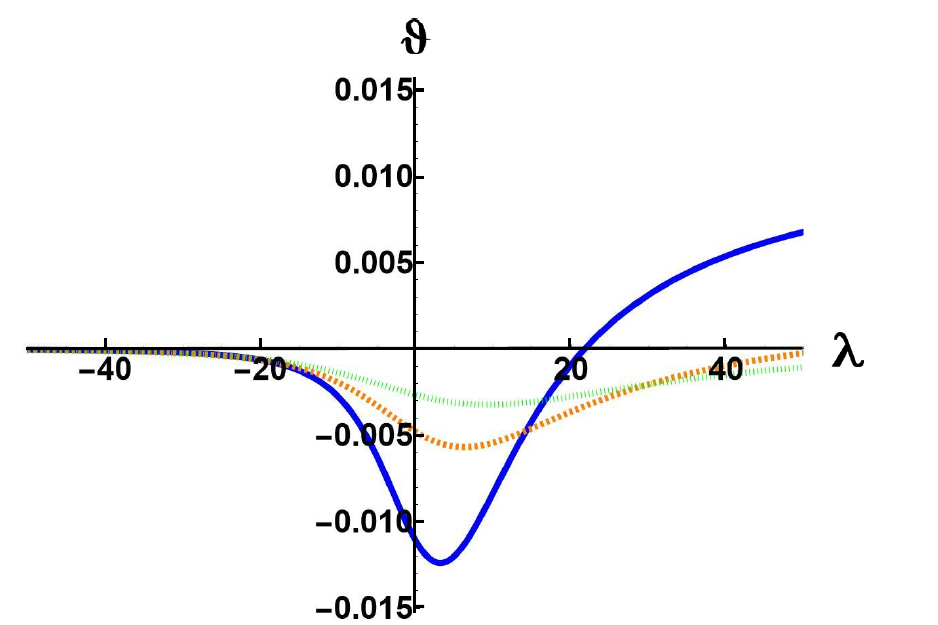}
\caption{\n{TM_l}A plot of $\vartheta(\lambda)$ for retrograde rays with fixed rotation parameter $\alpha=0.5$ and varying $\rho_m$=10 (solid), 15 (dashed), and 20 (dotted).}
\end{figure}
Let $\lambda_{in}$ be a large negative parameter. For the adopted value $\lambda_0=\lambda_m$, one has
the following initial conditions for the incoming rays
\be \n{TPics}
\rho_0=\lambda_{in}\hhh
\dot{\rho}_0=-1\hhh \vartheta_0=0\hhh
\dot{\vartheta}_0=0\, .
\ee
Solving \eqref{ddotr} and \eqref{ddotl} obtains a function $\vartheta=\vartheta(\lambda)$, which in the limit $\lambda\to \infty$ determines the asymptotic tilting angle for the outgoing trajectory of the null rays.

In Figs.~\ref{TP_l} and \ref{TM_l}, we see how $\vartheta(\lambda)$ behaves for some fixed value of $\alpha$ and varying values of $\rho_m$, for both prograde and retrograde rays, respectively.

In the asymptotic region where $\lambda\to \infty$, $\dot{\rho}\approx 1$, and equation \eqref{ddotl} reduces to
\be
\ddot{\vartheta}\approx -\dfrac{2\dot{\vartheta}}{\rho}\, .
\ee
This gives $\dot{\vartheta}\approx C/\rho^2$, and
\be
\vartheta\approx \vartheta^{\infty}-\dfrac{C}{\rho}\, .
\ee

Here $C$ is a constant of integration, and $\vartheta^{\infty}$ is the asymptotic value of the tilting angle.
By integrating equations \eqref{ddotr} and \eqref{ddotl}, one can find the asymptotic value of this tilting angle.

\begin{figure}[!hbt]
    \centering
\includegraphics[width=0.35\textwidth]{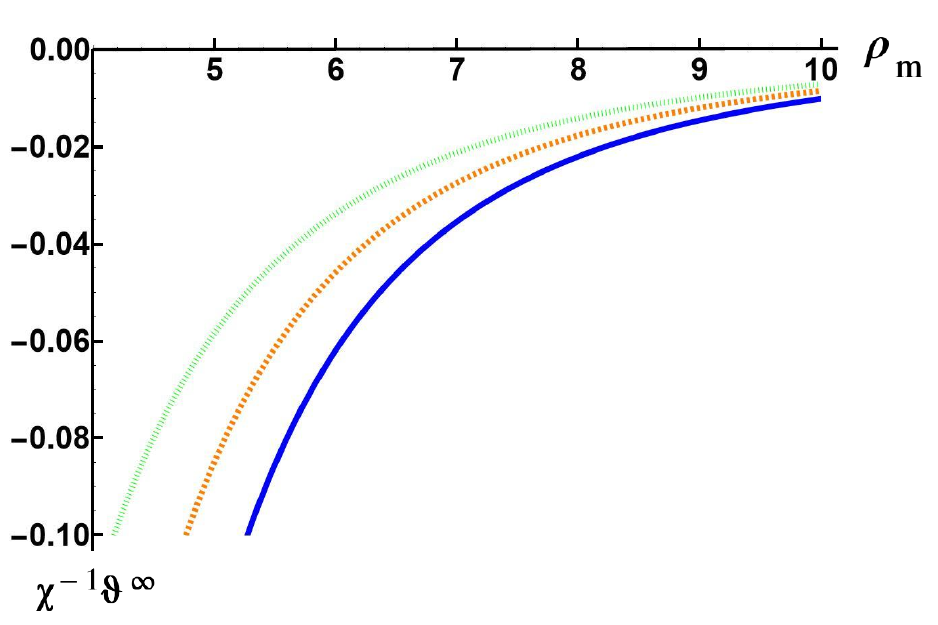}
\vspace{-6pt}
\caption{\n{TInfP_rm}A plot of $\chi^{-1}\,\vartheta^{\infty}$ as a function of $\rho_m$ for prograde photon scattering in the equatorial plane, with $\alpha =$ 0 (solid), 0.5 (dashed), 1 (dotted).}
\end{figure}

\begin{figure}[!hbt]
    \centering
\includegraphics[width=0.35\textwidth]{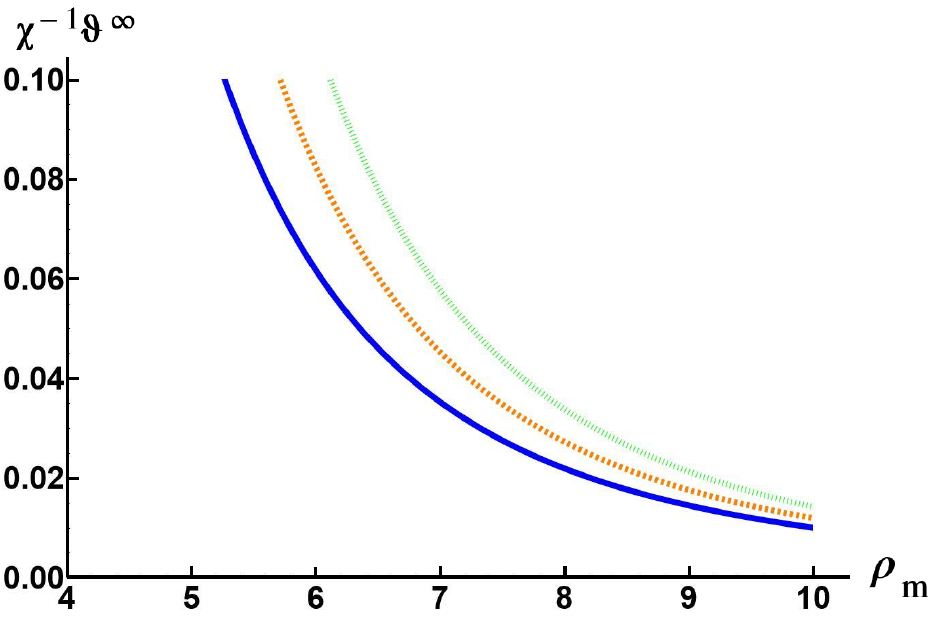}
\vspace{-6pt}
\caption{\n{TInfM_rm}A plot of $\chi^{-1}\,\vartheta^{\infty}$ as a function of $\rho_m$ for retrograde photon scattering in the equatorial plane, with $\alpha =$ 0 (solid), 0.5 (dashed), 1 (dotted).}
\end{figure}

We perform these calculations for the values $\alpha=0$, $\alpha=0.5$ and $\alpha=1$ of the dimensionless rotation parameter $\alpha$. For each of these values, we find $\vartheta^{\infty}$ as a function of the parameter $\rho_m$, which is the radial coordinate of the radial turning point of the ray trajectory. These plots for prograde and retrograde rays are presented in Figs.~\ref{TInfP_rm} and
\ref{TInfM_rm}, respectively.

For $\chi>0$  the asymptotic tilting angles $\vartheta^{\infty}$ for prograde and retrograde rays are correspondingly negative and positive. This means that after scattering, the prograge rays with positive helicity are reflected from the equatorial plane to the lower ``half-space", that is, to the ``south pole" black hole direction. The deflection of the positive helicity retrograde rays is in the opposite direction.

The plots \ref{TInfP_rm} and \ref{TInfM_rm} show that the absolute value of the tilting angle $\vartheta^{\infty}$ decreases with the growth of the parameter $\rho_m$. This is quite natural since the rays with larger $\rho_m$ values pass at larger distance from the black hole where the curvature is small. In Appendix~\ref{apxC}, the asymptotic values of the tilting angle $\vartheta^{\infty}$ for prograde and retrograde rays with large parameter $\rho_m$ are obtained. It it is shown that
\be \n{asvar}
\vartheta^{\infty}\approx \mp \dfrac{8\chi}{\rho^3_m}\,.
\ee
This expression does not depend on the rotation parameter $\alpha$. The dependence on $\alpha$ is present only in the higher-order terms of $\vartheta^{\infty}$ expansion over $1/\rho_m$.

Comparing the plots for prograde rays in Fig.~\ref{TInfP_rm}, one can see that for a given parameter $\rho_m$, the absolute value of the asymptotic tilting angle decreases with an increase in the rotation parameter $\alpha$. For retrograde rays the situation is different: the asymptotic tilting angle increases with an increase in the rotation parameter $\alpha$. This dependence of $\vartheta^{\infty}$ on the rotation can be explained as follows. The rotation dragging effect forces the prograde rays to ``move faster" near the black hole, so that they spend less time in the domain of high curvature. For retrograde rays, the effect is the opposite. For both types of rays, the asymptotic tilting angle formally grows infinitely when the parameter $\rho_m$ reaches its limiting value $\rho_m^{\pm}$ (see \eqref{rhopm}).

Figs.~\ref{TInfP_lz} and \ref{TInfM_lz} show a similar picture, except that they are dependent on the normalized impact parameter $\ell_z=L_z/M$ instead of the normalized minimum trajectory radius $\rho_m=r_m/M$.

\begin{figure}[!hbt]
    \centering
\includegraphics[width=0.35\textwidth]{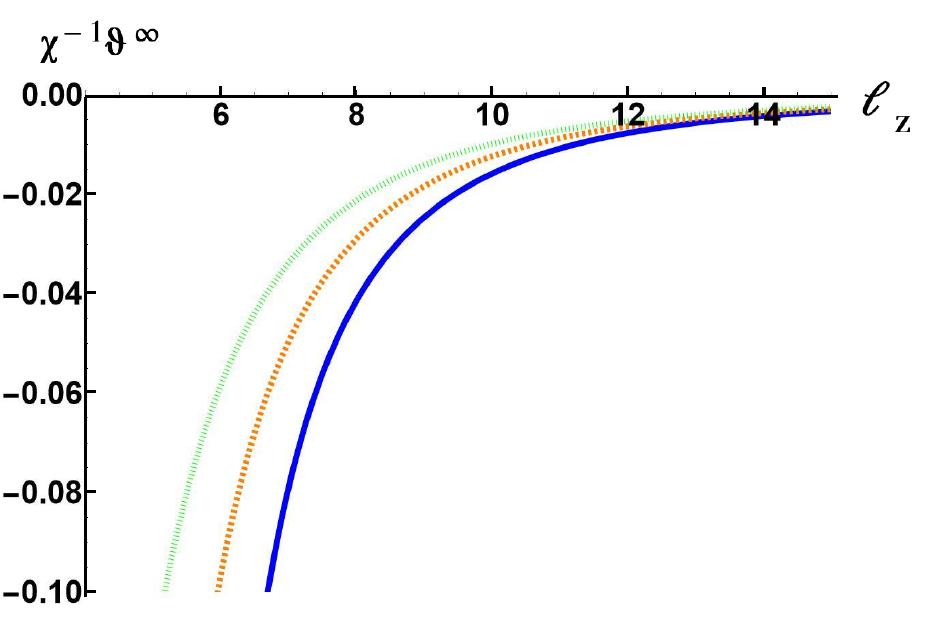}
\vspace{-6pt}
\caption{\n{TInfP_lz}A plot of $\chi^{-1}\,\vartheta^{\infty}$ as a function of $\ell_z$ for prograde photon scattering in the equatorial plane, with $\alpha =$ 0 (solid), 0.5 (dashed), 1 (dotted).}
\end{figure}

\begin{figure}[!hbt]
    \centering
\includegraphics[width=0.35\textwidth]{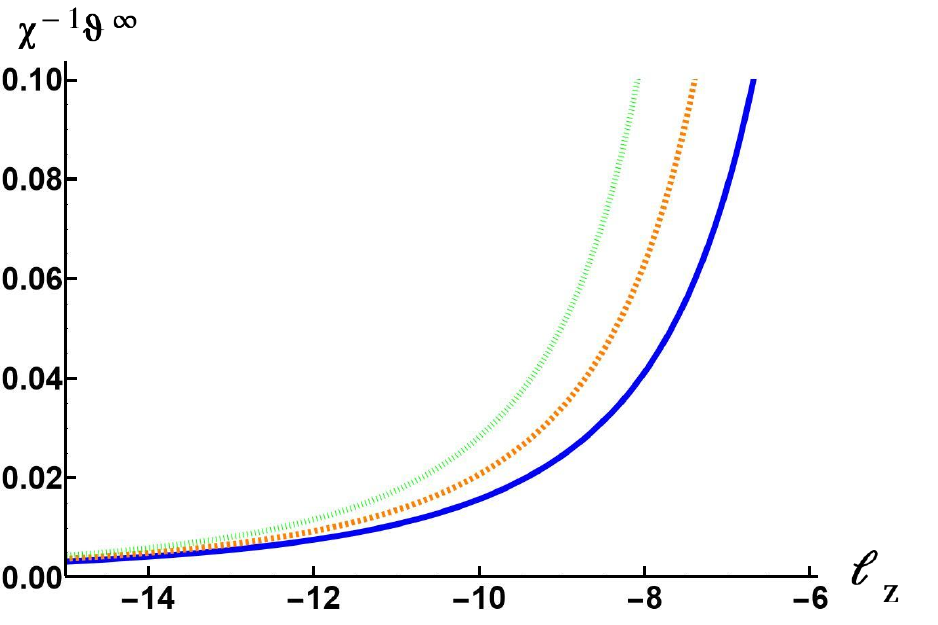}
\vspace{-6pt}
\caption{\n{TInfM_lz}A plot of $\chi^{-1}\,\vartheta^{\infty}$ as a function of $\ell_z$ for retrograde photon scattering in the equatorial plane, with $\alpha =$ 0 (solid), 0.5 (dashed), 1 (dotted).}
\end{figure}

\subsection{Tilting effect for non-rotating black holes}

It is instructive to consider the case of a non-rotating black hole, where the spinoptics equations are greatly simplified.
In dimensionless coordinates $(\tau,\rho,\theta,\phi)$, the Schwarzschild metric is
\be
ds^2=-\big(1-{2}/{\rho}\big)d\tau^2+
\dfrac{d\rho^2}{1-{2}/{\rho}}+\rho^2\big(d\theta^2+\sin^2\theta d\phi^2\big)\, .
\ee
It is evident that this metric is invariant under independent reflections of $\tau\to -\tau$, $\theta\to -\theta$ and $\phi\to -\phi$. The non-vanishing ZAMO projections of the Riemann tensor calculated on the equatorial plane $\theta=\pi/2$ are
\ba \n{Rproj}
R_{(\tau)(\rho)(\tau)(\rho)} = &-\frac{2}{\rho^3}\hhh R_{(\tau)(\theta)(\tau)(\theta)}=R_{(\tau)(\phi)(\tau)(\phi)}=\frac{1}{\rho^3}\,,\\
R_{(\rho)(\theta)(\rho)(\theta)}=& R_{(\rho)(\phi)(\rho)(\phi)}=-\frac{1}{\rho^3}\hhh
R_{(\theta)(\phi)(\theta)(\phi)}=\frac{2}{\rho^3}\,,
\ea
where now $\Delta=\rho^2-2\rho$. Note that we use the shorthand of parentheses to denote these ZAMO projections.  For example,
\be
R_{(\tau)(\rho)(\tau)(\rho)} = R_{\mu\nu\alpha\beta}E_{(\tau)}^{\mu}E_{(\rho)}^{\nu}E_{(\tau)}^{\alpha}E_{(\rho)}^{\beta}\,.
\ee
It is easy to see that these components are invariant under the reflection $\phi\to -\phi$.
Under this reflection, the Killing vector $\ts{\zeta}$ transforms as $\zeta^{\mu}\to -\zeta^{\mu}$. The tangent vector $\ts{l}$ to a null geodesic ray is also invariant under this reflection, provided that the corresponding angular momentum $\ell_z=\zeta^{\mu}l_{\mu}$ changes its sign, $\ell_z\to -\ell_z$. It easy to check that on the equatorial plane $\zeta^{\mu}h_{\mu\nu}=0$, hence $\zeta^{\mu}\tilde{e}_{3\mu}=0$, while
\be
\zeta^{\mu}{e}_{3\mu}=\Phi\ \zeta^{\mu}l_{\mu}=\Phi\ell_z=\lambda+
\lambda_m
\, .
\ee
The last equality is obtained by using \eqref{PPPP}. The above relation shows that under the reflection, we have $e_3^{\mu}\to -e_3^{\mu}$.
Using the above relations one concludes that
\be \n{psi2Sch}
\psi_2=R_{\mu\nu\alpha\beta}e_2^{\mu}l^{\nu}e_2^{\alpha}e_3^{\beta}
\ee
changes its sign under the reflection $\phi\to -\phi$. This means that the deflection angle $\vartheta^{\infty}$ for prograde null rays ($\ell_z>0$) is equal in magnitude but opposite in sign to the deflection angle for retrograde rays ($\ell_z>0$).

As another check, one can also calculate the asymptotic behavior for $\psi_2$ as $\rho\to\infty$.  To do this explicitly, we utilize equations \eqref{ASSIM} and \eqref{Rproj}, and substitute them into equation \eqref{psi2Sch}, only keeping terms of leading order in $\rho$.  One finds that the only curvature terms from \eqref{Rproj} that contribute are $R_{(\theta)(\tau)(\theta)(\tau)}$, $R_{(\theta)(\rho)(\theta)(\rho)}$, and $R_{(\theta)(\phi)(\theta)(\phi)}$, and we get
\be
\psi_{2}=\frac{3\epsilon_{\rho}\ell_z}{\rho^4}+ O(1/\rho^5)
\ee
This expression correctly reproduces the sign and value of $\psi_{2}$ at large distance from the black hole.

Let us note that in the absence of the ``driving force" $\psi_2$, equation \eqref{DDLL} reduces to the equation describing null geodesic deviation.
For discussion of this equation in the Schwarzschild geometry, see e.g. \cite{SCHW_GEOD} and references therein.

As we already mentioned, the spinoptics scattering problem in the Schwarzschild spacetime was considered in a recent paper \cite{Murk:2024qgj}. One can check that the acceleration vector $w^{\mu}$ of this paper coincides with the vector $\Upsilon^{\mu}$ evaluated for $\alpha=0$. The asymptotic expression for the tilting angle for prograde rays obtained in
\cite{Murk:2024qgj} for large $\rho_m$ (after writing it in the units and conventions adopted in the present paper) has the form
\be
\vartheta^{\infty}\approx -\dfrac{C\chi}{\rho_m}\, ,
\ee
where $C\approx 7.12$. This expression is similar to \eqref{asvar}, the only difference being that $C=8$ in \eqref{asvar}.

\section{Discussion} \label{s7}

In this paper, we discussed how the interaction of the spin of a massless field with the spacetime curvature modifies its propagates in curved spacetime\footnote{Let us mention that this effect is quite similar to the effect of the spin-curvature interaction of a spinning massive particle moving in a gravitational field (see e.g. \cite{SPIN_KERR_0,SPIN_KERR_1,SPIN_KERR_2} and references therein).}. To describe this effect, we used the spinoptics approach \cite{Oancea:2019pgm,Oancea:2020khc,
Frolov:2020uhn,Dahal:2022gop,Frolov:2024ebe,Frolov:2024qow}, which is a modification of the well-known geometric optics approach. In this approach, the problem of finding an approximate high-frequency solution of the wave equation is reduced to the study of dynamical equations for null rays and the null tetrads associated with them. In the leading zero order of a $1/\omega$ expansion, these equations describe geodesic null rays, while the associated tetrads are parallel propagated along them. In the spinoptics approximation, these equations are modified by the inclusion of first-order $1/\omega$ helicity-dependent corrections.

We focused on study of the spinoptics effect in the Kerr spacetime describing the gravitational field of a rotating black hole. In the Kerr metric, the geodesic equations and the parallel transport equations for the tetrad vectors can be solved in an explicit analytic form \cite{MARCK_1}. In the present paper, we used the method of solving the parallel transport equations developed in \cite{Kubiznak:2008zs}. (See also \cite{Dahal:2023ncl}).
The spinoptics effect is characterized by a small dimensionless parameter $\chi=\sigma/(M\omega)$.
To describe the interaction of the spin of the massless field with the spacetime curvature, we used the method of perturbations. Since the ``driving force" that describes this interaction is proportional to $\chi$, for the calculation of the driving force one can use unperturbed null tetrads associated with a corresponding null geodesic.

To describe a deviation of the ``accelerated" null ray from a geodesic one, we derived a ray deviation equation, which is a natural generalization of the deviation equation for null geodesics. We used this equation to study the propagation of null rays in the vicinity of the equatorial plane. We demonstrated that for the scattering of circularly polarized waves, the asymptotic planes of the orbits of incoming and outgoing null rays are different. We calculated the corresponding tilting angles and studied their dependence on the ray impact parameter and on the spin of the black hole.
For a non-rotating black hole, the result for the calculated tilting angle agrees with the results presented in \cite{Dahal:2021qel,Murk:2024qgj,Frolov:2024olb}.

As we already mentioned in the Introduction, a recent paper by Dahal \cite{Dahal:2023ncl} discussed some spinoptics effects in the Kerr metric. Its main result was the calculation of the tilting angle for scattering of null rays passing at large distance from the black hole. Our results are a far-reaching generalization of the results presented in \cite{Dahal:2023ncl}.
In particular, we discussed and compared the asymptotic tilting angles for the scattering of polarized light near the equatorial plane for prograde and retrograde rays. This allowed us to illustrate the dependence of the tilting angle on the rotation parameter of the black hole.

Let us mention that there exist other interesting spinoptical effects for the light and gravitational wave propagation near black holes. If a small-size object emits polarized radiation its image on the observer's screen evaluated in the spinoptics approximation, would be slightly shifted. This shift depends on the helicity, and for a given helicity, it also depends on the frequency of the radiation. As a result, instead of a point-like image of such an object, one observes a ``colored" strip. This ``rainbow effect" is quite interesting feature of spinoptics. Another spin-optical effect is an additional helicity- and frequency-dependent time delay for waves propagating near black holes.
Let us also mention that an interesting question is how the spinoptical effects can modify the shadow of a black hole.
We hope to address these questions in a separate paper.

\appendix

\section{Covariant components of the null tetrad vectors} \label{apxA}

For completeness, we give the covariant form of the vectors of the tetrad $(\ts{l},\tilde{\ts{n}},\ts{e}_2,\ts{\tilde{e}}_3)$ associated with a null geodesic in the Kerr metric. Written in the  dimensionless coordinates $(\tau,\rho,\theta,\phi)$ these vectors are
\ba
l_{\mu}=&\bigg(-1,\,\epsilon_{\rho}\,\frac{\sqrt{\CAL{R}}}{\Delta},\,\epsilon_{\theta}\sqrt{\Theta},\,\ell_z\bigg)\,,\\
\tilde{n}_{\mu}=&\frac{1}{2Q}\bigg(\Sigma-2(\rho^2+\alpha^2-\ell_{z}\alpha),\\
&\hspace{35pt}\Sigma\epsilon_{\rho}\sqrt{\CAL{R}}/\Delta,\,-\Sigma\epsilon_{\theta}\sqrt{\Theta},\\
&\hspace{35pt}(\rho^2+\alpha^2)(\alpha\sin^2\theta-\ell_{z})\\
&\hspace{35pt}+\alpha\sin^2\theta(\rho^2+\alpha^2-\ell_{z}\alpha)\bigg)\,,\\
e_{2\mu}=&\frac{1}{\Sigma\sqrt{Q}}\bigg(-\alpha\big(\cos\theta\epsilon_{\rho}\sqrt{\CAL{R}}-\rho\sin\theta \epsilon_{\theta}\sqrt{\Theta}\big),\\
&\hspace{35pt}\frac{\Sigma \alpha\cos\theta}{\Delta}(\rho^2+\alpha^2-\ell_{z}\alpha),\\
&\hspace{35pt}-\Sigma \rho\sin\theta\big(\alpha-\ell_{z}/\sin^2\theta\big),\\
&\hspace{35pt}\sin\theta\Big(\alpha^2\sin\theta\cos\theta\epsilon_{\rho}\sqrt{\CAL{R}}\\
&\hspace{35pt}-\rho(\rho^2+\alpha^2)\epsilon_{\theta}\sqrt{\Theta}\Big)\bigg)\,,\\
\tilde{e}_{3\mu}=&\frac{-1}{\Sigma\sqrt{Q}}\bigg(-\Big(\rho\epsilon_{\rho}\sqrt{\CAL{R}}+\alpha^2\sin\theta\cos\theta\epsilon_{\theta}\sqrt{\Theta}\Big),\\
&\hspace{35pt}+\frac{\Sigma}{\Delta}\Big(\rho(\rho^2+\alpha^2-\ell_{z}\alpha)\Big),\\
&\hspace{35pt}+\Sigma\Big(\alpha\sin\theta\cos\theta\big(\alpha-\ell_{z}/\sin^2\theta\big)\Big),\\
&\hspace{35pt}+\alpha\sin\theta\Big(\rho\sin\theta\epsilon_{\rho}\sqrt{\CAL{R}}\\
&\hspace{35pt}+(\rho^2+\alpha^2)\cos\theta\epsilon_{\theta}\sqrt{\Theta}\Big)\bigg)\,.
\ea

The expressions for the vectors of the parallel propagated tetrad $(\ts{l},\ts{{n}},\ts{e_2},\ts{{e}_3})$ can easily be obtained using relations \eqref{ee33} and \eqref{nnPP}.

Note that
\be
l_{[\mu,\nu]} = 0\,.
\ee
One can easily show that
\be
\begin{split}
&l_{\mu}=U_{,\mu}\, ,\\
&U=-\tau+\ell_z\phi +\int \dfrac{\epsilon_\rho\sqrt{\CAL{R}}d\rho}{\Delta}+\int \epsilon_{\theta} \sqrt{\Theta}d\theta\, .
\end{split}
\ee

Note that we can also write the vectors $(\ts{l},\ts{\tilde{n}},\ts{e_2},\ts{\tilde{e}_3})$ in terms of the ZAMO basis given in equation \eqref{EEEE}.  In the equatorial plane, this amounts to
\ba\n{TET_ZAM}
l^{\mu}=&\frac{A-2\rho\alpha\ell_z}{\rho\sqrt{\Delta A}}E_{(\tau)}^{\mu}+\epsilon_{\rho}\rho\sqrt{\frac{\CAL{B}}{\Delta}}E_{(\rho)}^{\mu}+\frac{\rho\ell_z}{\sqrt{A}}E_{(\phi)}^{\mu}\,,\\
\tilde{n}^{\mu}=&\frac{\rho\big((\rho^2+\alpha^2-\alpha\ell_z)^2+\alpha^2\Delta-\alpha\ell_z(\alpha\ell_z-2\rho)\big)}{2\sqrt{A\Delta}(\ell_z-\alpha)^2}E_{(\tau)}^{\mu}\\
&+\frac{\rho\big(2\alpha(\rho^2+\alpha^2-\alpha\ell_z)-\ell_z\rho^2\big)}{2(\ell_z-\alpha)^2\sqrt{A}}E_{(\phi)}^{\mu}\\
&+\frac{\epsilon_{\rho}\rho^3\sqrt{\CAL{B}}}{2(\ell_z-\alpha)^2\sqrt{\Delta}}E_{(\rho)}^{\mu}\,,\\
e_{2}^{\mu}=&E_{(\theta)}^{\mu}\,,\\
\tilde{e}_{3}^{\mu}=&\frac{\epsilon_{\rho}\rho^2(
\rho^2+\alpha^2)\sqrt{\CAL{B}}}{(\ell_z-\alpha)\sqrt{A\Delta}}E_{(\tau)}^{\mu}-\frac{\rho^2-\alpha(\ell_z-\alpha)}{(\ell_z-\alpha)\sqrt{\Delta}}E_{(\rho)}^{\mu}\\
&-\frac{\epsilon_{\rho}\rho^{2}\alpha\sqrt{\CAL{B}}}{(\ell_z-\alpha)\sqrt{A}}E_{(\phi)}^{\mu}\,.
\ea

\section{Non-geodesic null ray's congruence
} \label{apxB}

Consider the congruence of null rays $x^{\mu}=x^{\mu}(\lambda,y^i)$, where $\lambda$ is a parameter along a null ray, and $y^i$, with $i=1,2,3$, are parameters that ``enumerate" the rays.
Denote by $l^{\mu}$ the tangent vectors to the rays
\be
l^{\mu}=\dfrac{\pa x^{\mu}}{\pa\lambda}\, .
\ee
If a ray obeys the equation
\be\n{geod}
Dl^{\mu}=A l^{\mu}\hh D=l^{\alpha}\nabla_{\alpha}\, ,
\ee
then its integral line is a geodesic.

Let us change the parametrization
\be
\lambda\to \tilde{\lambda}
\ee
and denote $b=d\tilde{\lambda}/d\lambda$.
Then
\be
l^{\mu}=b\tilde{l}^{\mu}\, ,
\ee
and equation \eqref{geod} takes the form
\be
\tilde{D}\tilde{l}^{\mu}=\dfrac{1}{b}(A-\tilde{D}b)\tilde{l}^{\mu}
\hh
\tilde{D}=\tilde{l}^{\alpha}\nabla_{\alpha}\, .
\ee
Taking $b$ to be a solution of the equation
\be
\tilde{D}b=A\, ,
\ee
one gets
\be
\tilde{D}\tilde{l}^{\mu}=0\, .
\ee
We call such a choice of the parameter $\lambda$ along null rays ``canonical".

Let us consider a congruence of non-geodesic null rays, and denote by $(\ts{l},\hat{\ts{n}},\hat{\ts{m}},\bar{\hat{\ts{m}}})$ a complex null tetrad along the null rays that satisfies the following normalization conditions
\be \n{NORM}
l_{\mu}\hat{n}^{\mu}=-1\hh
\hat{m}_{\mu}\bar{\hat{m}}^{\mu}=1\, .
\ee
The other scalar products between the tetrad's vectors vanish.
For a general congruence of non-geodesic null-rays, one can write
\be \n{hatl}
Dl^{\mu}=\hat{w}_0 l^{\mu}+\bar{\hat{\kappa}}\hat{m}^{\mu}+\hat{\kappa}\bar{\hat{m}}^{\mu}\, .
\ee
The term containing $\hat{n}^{\mu}$ in the right-hand side of this relation vanishes because of the normalization conditions given in \eqref{NORM}. The coefficient $\hat{w}_0$ is real, while $\hat{\kappa}$ is complex.

Similarly, using the conditions in \eqref{NORM}, one can write the following relation
\be \n{hatm}
D\hat{m}^{\mu}=\hat{\mu}_0 l^{\mu}+i\hat{\mu}\hat{m}^{\mu}+\hat{\kappa}\hat{n}^{\mu}\, .
\ee
Here $\hat{\mu}$ is real and $\hat{\mu}_0$ is complex.

There exists a set of transformations that preserve the direction of $\ts{l}$ and the normalization conditions \eqref{NORM}, which can be used to simplify relations \eqref{hatl} and \eqref{hatm}. These transformations are
\begin{enumerate}
\n{TRANS}
\item $m^{\mu}\to m^{\mu}+\beta l^{\mu}\hh \bar{m}^{\mu}\to \bar{m}^{\mu}+\bar{\beta} l^{\mu}$ \, ,\n{11}\\
$n^{\mu}\to n^{\mu}+\beta \bar{m}^{\mu}+
\bar{\beta} {m}^{\mu}+\beta \bar{\beta}l^{\mu}
$\, ,
\item $m^{\mu}\to e^{i\varphi}m^{\mu}\hh
\bar{m}^{\mu}\to e^{-i\varphi}\bar{m}^{\mu} \, ,$\n{22}
\item $ l^{\mu}\to \alpha l^{\mu}\hh
n^{\mu}\to \alpha^{-1} n^{\mu} \, .$\n{33}
\end{enumerate}
Here $\alpha$ and $\varphi$ are real and $\beta$ is complex.

We first apply transformation \eqref{11}, and write
\be
\begin{split}
\hat{m}^{\mu}&= \check{m}^{\mu}+\beta l^{\mu}\hh \bar{\hat{m}}^{\mu}=
\bar{\check{m}}^{\mu}+\bar{\beta} l^{\mu}\, ,\\
\hat{n}^{\mu}&=\check{n}^{\mu}+\beta \bar{\check{m}}^{\mu}+
\bar{\beta} \check{m}^{\mu}+\beta \bar{\beta}l^{\mu}\,.
\end{split}
\ee
As a result of this transformations the equations \eqref{hatl} and \eqref{hatm} take the form
\be\n{check}
\begin{split}
Dl^{\mu}&=\check{w}_0l^{\mu}+\bar{\check{\kappa}}\check{m}^{\mu}+\check{\kappa}\bar{\check{m}}^{\mu}\, ,\\
D\check{m}^{\mu}&=\check{\mu}_0 l^{\mu}+i\check{\mu}\check{m}^{\mu}+\check{\kappa}\check{n}^{\mu}\, .
\end{split}
\ee
The coefficients on the right-hand side of these relations are
\ba
\check{w}_0=&\hat{w}_0+\bar{\hat{\kappa}}\beta+\hat{\kappa}\bar{\beta}\,,\\
\check{\mu}_0=&\hat{\mu}_0+i\hat{\mu}\beta-\beta\hat{w}_0-\bar{\hat{\kappa}}\beta^2-D\beta\,,\\
\check{\mu}=&\hat{\mu}-\hat{\kappa}\bar{\beta}+\bar{\hat{\kappa}}\beta\,,\\
\check{\kappa}=&\hat{\kappa}\,.
\ea
One can see that the coefficient $\check{\mu}_0$ can be put equal to zero if $\beta$ satisfies the following equation
\be
D\beta=\hat{\mu}_0+i\hat{\mu}\beta-\beta\hat{w}_0-\bar{\hat{\kappa}}\beta^2\,.
\ee
We use this choice of $\beta$ and write the second equation of \eqref{check} as follows
\be \n{NOMU_0}
D\check{m}^{\mu}=i\check{\mu}\check{m}^{\mu}+\check{\kappa}\check{n}^{\mu}\, .
\ee

We now apply the second transformation \eqref{22} and write
\be
\check{m}^{\mu}=e^{i\varphi}m^{\mu}\, .
\ee
If we choose the phase parameter $\varphi$ to satisfy the equation
\be
D\varphi=\check{\mu}\, ,
\ee
then we can present \eqref{NOMU_0} as follows
\be
D{m}^{\mu}={\kappa}n^{\mu}\, ,
\ee
where $\kappa=\check{\kappa}e^{-i\varphi}$ and $n^{\mu}=\check{n}^{\mu}$. In these notations, the first equation in \eqref{check} becomes
\be \n{DLW}
Dl^{\mu}={w}_0 l^{\mu}+\bar{{\kappa}}{m}^{\mu}+{\kappa}\bar{{m}}^{\mu}\, .
\ee
where $w_0=\check{w}_0$.

Let us summarize. Using transformations 1 and 2 from List~\ref{TRANS} one can find such complex null tetrad $(\ts{l},\ts{n},\ts{m},\bar{\ts{m}})$ in which the equations of motion of the null tetrad propagated along the null rays $\ts{l}$ take the form \cite{Frolov:2020uhn}
\be \n{EQL}
\begin{split}
& Dl^{\mu}=w^{\mu}\hhh
w^{\mu} =w_0 l^{\mu}+
\bar{\kappa} m^{\mu}+{\kappa}\bar{m}^{\mu}\, ,\\
&Dn^{\mu} = w_0 n^{\mu}\hh Dm^{\mu}=\kappa n^{\mu}\,  .
\end{split}
\ee
One can also use transformation \eqref{33} to cancel the parameter $w_0$ from these equations.

\section{Null ray deviation equation} \label{DEV}

In the presence of the ``driving force" $\chi \Upsilon^{\mu}$, the null rays propagate slightly differently than geodesic rays with $\chi=0$. To derive an equation describing their deviation from a corresponding null geodesic, we proceed as follows. Let us denote the tangent vector to an ``accelerated" null ray by $l^{\mu}$, and the corresponding null geodesic with $\chi=0$ by $l_0^{\mu}$.
Let $O$ be a point on a null geodesic.
We consider a one-parameter family of null rays for which the tangent vector $\ts{l}$ at $O$ coincides with $\ts{l}_0$,
and which have an acceleration equal to $v \Upsilon^{\mu}$. We assume that $v\in [0,\chi]$ (see Fig.~\ref{lpFig}).

\begin{figure}[!hbt]
    \centering
\includegraphics[width=0.45\textwidth]{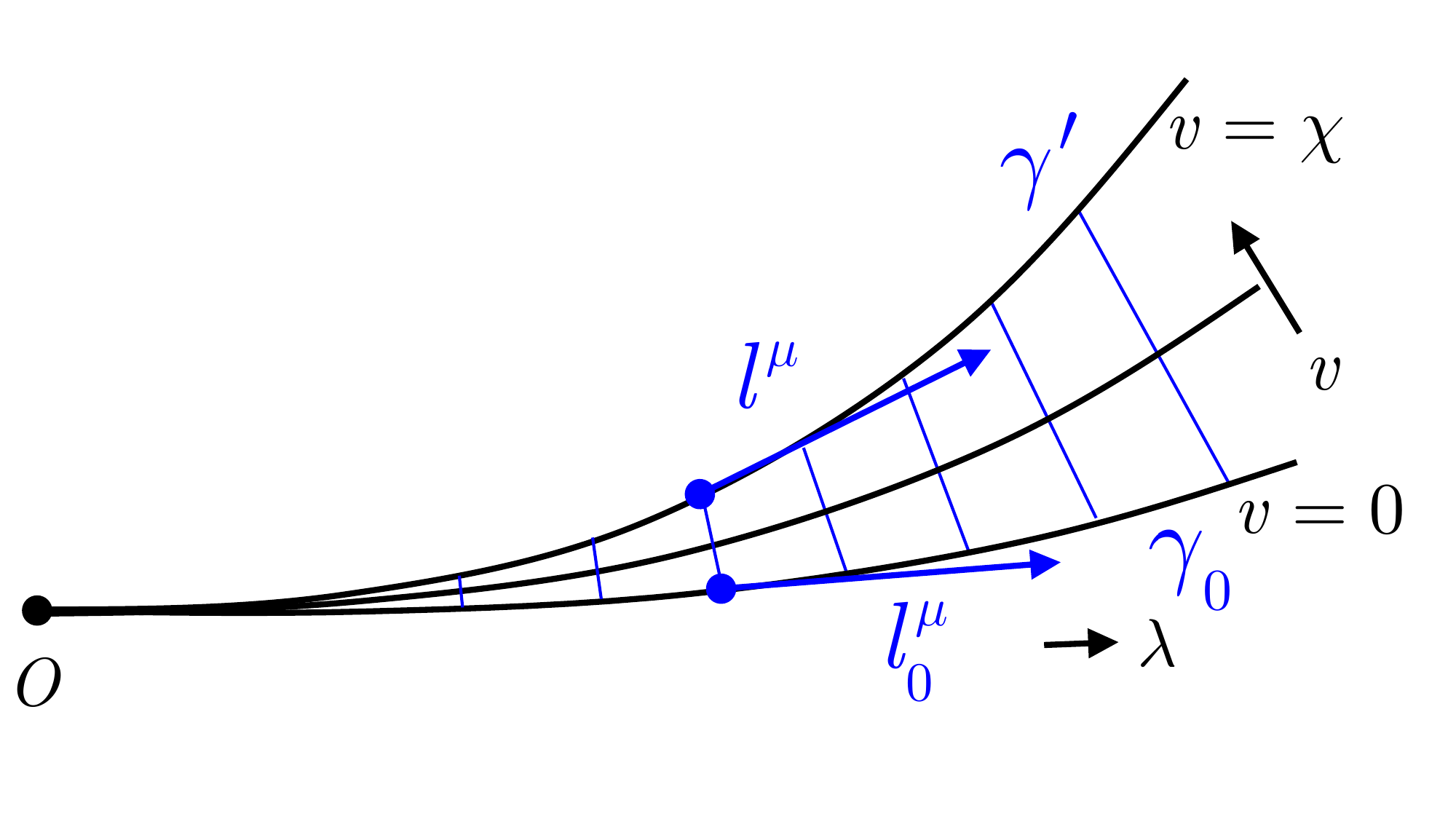}
%\vspace{-3cm}
\caption{\n{lpFig}Congruence of null rays spanned by perturbing a geodesic null ray with a small acceleration $v\Upsilon^{\mu}$, governed by small parameter $v$.}
\end{figure}

This set of rays covers a two-dimensional surface
\be
x^{\mu}=x^{\mu}(\lambda,v)\,,
\ee
where $\lambda$ is a canonical parameter along a null ray, so that
\be
l^{\mu}=\dfrac{\pa x^{\mu}}{\pa \lambda}\, .
\ee
Denote
\be
p^{\mu}=\dfrac{\pa x^{\mu}}{\pa v}\, .
\ee
Then the deflection $\delta x^{\mu}(\lambda)$ of a point with parameter $\lambda$ along the null ray with acceleration $v\Upsilon^{\mu}$
from a point with the same $\lambda$ on the geodesic ray ($v=0$) is
\be \n{defl}
\delta x^{\mu}(\lambda)=v p^{\mu}\, .
\ee
A schematic drawing of the setup is shown in Fig.~\ref{lpFig}. Since partial derivatives with respect to $\lambda$ and $v$ commute, one has
\be
l^{\alpha}p^{\mu}_{\, ,\alpha}=p^{\alpha}l^{\mu}_{\, ,\alpha}\, .
\ee
This relation remains valid if the partial derivatives are changed by the covariant derivatives
\be \n{PPLL}
\nabla_p l^{\mu}=\nabla_{l}p^{\mu}\, .
\ee
For a pair of commuting vectors $\ts{l}$ and $\ts{p}$,
one has
\be
\nabla_{l}\nabla_p l^{\mu}-\nabla_{p}\nabla_l l^{\mu}=R^{\mu}_{\alpha\beta\gamma}l^{\alpha}l^{\beta}p^{\gamma}\, .
\ee
This relation directly follows from the definition of the Riemann tensor.
Using \eqref{PPLL}, one can write the first term in the left hand side as $\nabla_{l}\nabla_p l^{\mu}=\nabla_{l}\nabla_{l} p^{\mu}$, while the second term is
\be
\nabla_{p}\nabla_{l} l^{\mu}=p^{\alpha}(v \Upsilon^{\mu})_{;\alpha}=\Upsilon^{\mu}+v p^{\alpha} \Upsilon^{\mu}_{;\alpha}\, .
\ee

Using definition \eqref{defl} for the displacement vector $\delta x^{\mu}$, and putting $v=\chi$,  one gets
\be \n{DDLL}
D_0^2 \delta x^{\mu}-R^{\mu}_{\alpha\beta\gamma}l_0^{\alpha}l_0^{\beta}\delta x^{\gamma}=\chi \Upsilon^{\mu}\, .
\ee
In this relation we omitted any terms that were higher than first order  in $\chi$. We also denote $D_0=\nabla_{l_0}$, and identify $\delta x^{\mu}$ as a displacement of the null ray with acceleration $\chi \Upsilon^{\mu}$ from a corresponding null geodesic. Let us note that for $\Upsilon^{\mu}=0$, equation \eqref{DDLL} is simply the standard deviation equation for null geodesics.

\section{Scattering with large impact parameter} \label{apxC}

Here we discuss the scattering of rays with large impact parameter $\ell_z$ in the vicinity of the equatorial plane of the Kerr black hole. For large $\ell_z$, the dimensionless radius of the radial turning point $\rho_m$ is also large. To find the solutions of \eqref{XX2} in this regime, we introduce the following new variables
\be \n{rhom}
\begin{split}
\hat{\rho}&=\rho/\rho_m\, ,\ \hat{\ell}_z=\ell_z/\rho_m\, ,\
\hat{\lambda}=\lambda/\rho_m\, ,\\
Y&=-\rho_m \CAL{X}_2/(3\chi \hat{\ell}_z)\,.
\end{split}
\ee
Then \eqref{XX2} takes the form
\be\n{YYY}
\dfrac{d^2 Y}{d\hat{\lambda}^2} +\dfrac{3 \hat{\ell}_z^2}{\hat{\rho}^5}\frac{Y}{\rho_m}=  \dfrac{\hat{\lambda}+\hat{\lambda}_m}{\hat{\rho}^5}\, .
\ee
Using equations \eqref{rmlarge} and \eqref{LMR}, one has
\be
\hat{\ell}_z\approx \pm (1+1/\rho_m)\hh
\hat{\lambda}_m\approx 1/\rho_m\, .
\ee

Let us write  a solution of \eqref{YYY} in the form
\be \n{Y01}
Y=Y_0+\dfrac{1}{\rho_m}Y_1+...\, .
\ee
Then
\be\n{YY0}
\dfrac{d^2 Y_0}{d\hat{\lambda}^2}= \dfrac{\hat{\lambda}}{\hat{\rho}^5}\, .
\ee
For large $\rho_m$, one has can find a relation between $\hat{\lambda}$ and $\hat{\rho}$ by using Eq.~\eqref{Brm}. We put $\hat{\rho}=\cosh\mu$, and therefore we find that
\be
\hat{\lambda}\approx\sinh\mu+(1/\rho_m)\tanh(\mu/2)\, .
\ee
In the leading order $\hat{\lambda}\approx\sinh\mu$
and so \eqref{YY0} takes the form
\be
\dfrac{d^2 Y_0}{d\mu^2}
-\dfrac{\sinh\mu}{\cosh\mu}\dfrac{d Y_0}{d\mu}
= \dfrac{\sinh\mu}{\cosh^3\mu}\, .
\ee
A solution of this equation is
\be
Y_0=-\frac{1}{3}\tanh\mu+c_1\sinh\mu+c_2\, ,
\ee
where $c_1$ and $c_2$ are integration constants. The boundary condition $Y_0(\mu=-\infty)=0$ implies that
\be \n{Y0SOL}
Y_0=-\frac{1}{3}(\tanh\mu+1)\, .
\ee
Then for $\mu\to \infty$ one has $Y_0=-2/3$. This according to Eq.~\eqref{rhom}, at zeroth order
\be
\vartheta=-\dfrac{3\chi\hat{\ell}_z}{\rho_m^2} \dfrac{Y_0}{\cosh\mu}\, .
\ee
This means that for $Y_0$ the tilting angle is $\vartheta^{\infty}=0$.

Substituting the obtained solution \eqref{Y0SOL} and now taking \eqref{YYY} to first order in the perturbation \eqref{Y01}, one obtains a differential equation for $Y_1$
\ba
\frac{(\sinh\mu+\cosh\mu+1)^2}{\cosh^{5}\mu(\cosh\mu+1)}=&\frac{d}{d\mu}\Big[\frac{1}{\cosh{\mu}}\frac{dY_1}{d\mu}\\
&+\frac{1}{3\cosh^4{\mu}(\cosh\mu+1)}\Big]\,.
\ea
One can solve this equation to obtain $Y_1$ explicitly. As before, the condition that $Y_1(\mu=-\infty)=0$ restricts our two constants of integration, and we have
\ba
\\
Y_1 = &\frac{4}{3}(\cosh\mu+\sinh\mu)-\frac{1}{4}\tan^{-1}(\sinh\mu)\\
&-\tan^{-1}(\tanh(\mu/2))+\frac{1}{3}\tanh(\mu/2)\\
&+\frac{\tanh(\mu/2)}{\tanh^2(\mu/2)+1}-\frac{2\tanh(\mu/2)}{3\big(\tanh^2(\mu/2)+1\big)^2}\\
&-\frac{1}{4}\frac{\sinh\mu}{\cosh^{2}\mu}-\frac{2}{3}\frac{1}{\cosh\mu}+\frac{2}{3}-\frac{5\pi}{8}\,.
\ea
One can show that only the first term in parentheses on the right-hand side of this expression after its division by $\cosh\mu$ remains finite as $\mu\to\infty$, and one has
${Y_1}/{\cosh\mu}\to 8/3$.
This allows us to find $\vartheta^{\infty}$, which has the form
\be\n{TILT}
\vartheta^{\infty}\approx \mp \dfrac{8\chi}{\rho^3_m}\,
\ee

%%%%%%%%%%%%%%%%%%%%%%%%%%%%%%%%%%%%%%%%%

\section*{Acknowledgements}
The authors thank Prof. Don Page for stimulating discussions and useful comments.
This work was supported by the Natural Sciences and
Engineering Research Council of Canada. One of the authors (V.F.) is also grateful to the Killam Trust for its financial support.

%%%%%%%%%%%%%%%%%%%%%%%%%%%%%%%%%%%%%%%

\bibliography{1_main}

\end{document}